\documentclass[aps, prd, amsfonts, a4paper,floatfix,twocolumn]{revtex4}

\usepackage{graphicx}
\usepackage{amsmath,amsfonts}
\usepackage{epsfig,color}

\begin{document}
\newcommand{\be}{\begin{eqnarray}}
\newcommand{\ee}{\end{eqnarray}}
\newcommand\del{\partial}
\newcommand\nn{\nonumber}
\newcommand{\Tr}{{\rm Tr}}
\newcommand{\Str}{{\rm Trg}}
\newcommand{\mat}{\left ( \begin{array}{cc}}
\newcommand{\emat}{\end{array} \right )}
\newcommand{\vect}{\left ( \begin{array}{c}}
\newcommand{\evect}{\end{array} \right )}
\newcommand{\tr}{{\rm Tr}}
\newcommand{\hm}{\hat m}
\newcommand{\ha}{\hat a}
\newcommand{\hz}{\hat z}
\newcommand{\hx}{\hat x}
\newcommand{\hl}{\hat \lambda}
\newcommand{\tm}{\tilde{m}}
\newcommand{\ta}{\tilde{a}}
\newcommand{\tz}{\tilde{z}}
\newcommand{\tx}{\tilde{x}}
\definecolor{red}{rgb}{1.00, 0.00, 0.00}
\newcommand{\rd}{\color{red}}
\definecolor{blue}{rgb}{0.00, 0.00, 1.00}
\definecolor{green}{rgb}{0.10, 1.00, .10}
\newcommand{\blu}{\color{blue}}
\newcommand{\green}{\color{green}}



\title{Finite-Volume Scaling of the Wilson-Dirac Operator Spectrum}

\author{P.H. Damgaard}
\affiliation{Niels Bohr International Academy and Discovery Center, Niels Bohr Institute, University of Copenhagen,
  Blegdamsvej 17, DK-2100, Copenhagen {\O}, Denmark} 

\author{U.M. Heller}
\affiliation{American Physical Society, One Research Road, Ridge, NY 11961, USA}

\author{K. Splittorff}
\affiliation{Discovery Center, Niels Bohr Institute, University of Copenhagen, Blegdamsvej 17, DK-2100, Copenhagen
  {\O}, Denmark}

\date   {\today}
\begin  {abstract}
The microscopic spectral density of the Hermitian Wilson-Dirac
operator is computed numerically in quenched lattice QCD. We
demonstrate that the results given for fixed index of the Wilson-Dirac 
operator can be matched by the predictions from Wilson chiral 
perturbation theory. We test successfully the finite volume and 
the mass scaling predicted by Wilson chiral perturbation theory  
at fixed lattice spacing. 
\end{abstract}
\maketitle

\section{Introduction}
\label{sec:Intro}

For a long time, the Wilson construction of Dirac fermions on the lattice was
somewhat overshadowed by other formulations with either explicit chiral symmetry
or remnants thereof. Recently, with the help of improved simulation techniques
and more powerful computers, numerical simulations with Wilson fermions have
picked up momentum again. The initial concern about the explicit breaking of chiral 
symmetry in this formulation can be dealt with in precisely the way originally
envisaged: by simply pushing simulations into the regime of very light fermions,
even to the physical point of almost massless $u$ and $d$ quarks. This requires
tight control on the smallest eigenvalues of the associated Wilson-Dirac
operator $D_W$. If the lattice spacing is not small enough, the gap associated
with the Hermitian Wilson-Dirac operator $D_5 \equiv \gamma_5(D_W+m)$, where $m$
is the quark mass in continuum language, shrinks due to small eigenvalues populating 
the region between $-m$ and $m$. 
This can cause severe numerical instabilities in the simulations. Studies of
the spectral gap and its potential disappearance have indicated that simulations
may be continued to the physical point even at realistic lattice spacings
of present-day computational power \cite{Luscher}.

Sparked by this renewed interest in simulations with Wilson fermions on the lattice, 
also work on analytical approaches to understanding the chiral limit of these fermions
has intensified. As is always the case, it is advantageous to phrase the
discussion in terms of the pertinent low-energy effective field theory, Wilson chiral
perturbation theory (WCPT) \cite{Sharpe}. Within this framework Sharpe \cite{Sharpe1} was 
the first to consider
the effects of finite lattice spacing, $a$, on the spectrum of the Hermitian Wilson-Dirac
operator. More recently, a complete description of the microscopic details of
the Wilson-Dirac operator spectrum up to and including $a^2$ corrections has been
presented for both quenched \cite{DSV,ADSV,AN,KVZ} and unquenched cases \cite{ADSVNf1,SVdyn}.
Here we will test these recent predictions against quenched lattice data.

The microscopic description works, at leading order in the chiral counting rule
of the finite-volume $\epsilon$-regime, at two equivalent levels: (i) a 
Wilson chiral random matrix theory (WRMT) and (ii) the zero-mode sector of WCPT to order $a^2$. 
Beyond leading order only the approach based on WCPT survives.

The $\gamma_5$-Hermiticity of the Wilson-Dirac operator,
\be
D_W^\dagger = \gamma_5 D_W \gamma_5,
\label{g5Herm}
\ee
together with the explicit form of $D_W$ tells us that the eigenvalues of $D_W$  
come in complex conjugate pairs \cite{Itoh}, a feeble remnant of chiral symmetry. 
In addition, $\gamma_5$-Hermiticity of $D_W$ ensures that the Hermitian 
Wilson-Dirac operator,
\be
\label{D5def}
D_5\equiv\gamma_5(D_W+m),
\ee 
as the name suggests, is Hermitian. While the eigenvalues of $D_W$ can be computed by,
say, the Arnoldi algorithm, it is clearly advantageous to focus instead on
the Hermitian Wilson-Dirac operator $D_5$, for which fast algorithms exist that
provide an ascending sequence of lowest eigenvalues. 

Here we test the predictions for the quenched microscopic spectral 
density of $D_5$ that were presented in \cite{DSV,ADSV} against lattice data. 
All results of \cite{DSV,ADSV} are given for fixed index of the Wilson-Dirac 
operator. This index $\nu$ is defined for a given gauge field configuration by 
\be
\label{defIndex}
\nu ~\equiv~ {\sum_{k}}' {\rm sign} (\langle k|\gamma_5|k\rangle).
\ee
Here, $|k\rangle$ denotes the $k$'th eigenstate of the Wilson-Dirac
operator. Only those eigenvectors which correspond to a real eigenvalue
of the Wilson-Dirac operator have a nonvanishing expectation value of
$\gamma_5$ \cite{Itoh} and hence contribute to the index. 

In the microscopic limit the real eigenvalues of the Wilson-Dirac
operator are predicted to be located in regions with width of order $1/V$
\cite{DSV,ADSV,KVZ}, where $V$ is the four volume, and we choose only to
include the physical/leftmost of these regions in the above definition of
the index. This is indicated by
the prime on the sum in Eq.~(\ref{defIndex}).

The predictions \cite{DSV,ADSV} for the quenched microscopic
spectral density of $D_5$ are based on the $\epsilon$-regime of 
Wilson chiral perturbation theory. At leading order in the $\epsilon$-regime 
of WCPT three additional low energy constants, $W_6$,
$W_7$ and $W_8$ parametrize the leading corrections due to the nonzero
lattice spacing (order $a^2$ corrections) \cite{Sharpe}. 
There are large-$N_c$ arguments ($N_c$ is the number of colors)
to the effect that the Wilson low-energy constants $W_6$ and $W_7$ may be small
\cite{Sharpe1}. For the case $W_6=W_7=0$ it was argued in \cite{DSV,ADSV}
that only $W_8>0$ correctly describes lattice QCD with a
nonnormal and $\gamma_5$-Hermitian Dirac operator.
In agreement with this we will demonstrate here that the 
quenched microscopic eigenvalue density of $D_5$ obtained on the lattice  
can be matched by the prediction from WCPT with the low energy parameters 
$W_6=0$, $W_7=0$ and $W_8>0$.   

We identify the effects of nonzero $W_6$ and $W_7$ in the spectrum of $D_5$, 
and conclude that they do not change the above conclusion. 

Finally, we explicitly compute the real eigenvalues of $D_W$ and
compare to the predictions.

Wilson chiral perturbation theory has also been considered in the 
finite-volume scaling limit where the $a^2$-terms can be pulled down 
from the action and treated as small perturbations \cite{Shindler}. 
The analytic constraint on $W_8$ that we discussed above is not obvious
in such a scaling limit.

Our paper is organized as follows. In the section just below,
we first briefly recall those results from WCPT in the $\epsilon$-regime 
that are most relevant for the present lattice study. Section \ref{sec:WRMT} 
gives the definition of Wilson chiral random matrix theory. In section 
\ref{sec:lat} we give details of the lattice
simulations we have performed. We compare our lattice data
to the predictions from Wilson chiral perturbation theory in section 
\ref{sec:results}. Section \ref{sec:conc} contains our conclusions.

\section{Predictions for the microscopic spectrum}
\label{sec:predictions}

Here we recall the predictions for the spectral density of $D_5$, 
\be
\rho_5(\lambda^5,m;a) = \left \langle \sum_k \delta(\lambda^5_k-\lambda^5) \right \rangle
\label{rho5def} .
\ee
The eigenvalues $\lambda^5$, defined by
\be
D_5\psi_k = \lambda^5_k \psi_k
\ee
are considered in the microscopic limit where $V\to\infty$ while  
\be
mV, \quad \lambda^5_k V, \quad \quad a^2V   
\ee
are kept fixed. 

The microscopic spectral density follows from WCPT. To ${\cal O}(a^2)$ 
accuracy this effective theory was written down in \cite{Sharpe}, expanding 
in all operators that contribute to
this order in the Szymanzik scheme. To order $a^2$ the zeromode partition function
at fixed index $\nu$  can be obtained \cite{DSV} by
decomposing the momentum zeromode partition function according to
\be
Z_{N_f}(m,\theta; a) \equiv \sum_{\nu=-\infty}^\infty e^{ i\nu\theta} Z_{N_f}^\nu(m;a). 
\label{decomb}
\ee
In this $\epsilon$-regime, the partition function reduces to a unitary matrix 
integral
\be
Z_{N_f}^\nu(m,z;a) =   \int_{U(N_f)} \hspace{-1mm} d U \ {\det}^\nu U
~e^{S[U]}, \label{Zfull}
\ee
where the action $S[U]$ for degenerate quark masses is given by
\be\label{lfull}
S & = & \frac{m}{2}\Sigma V{\rm Tr}(U+U^\dagger)+
\frac{z}{2}\Sigma V{\rm Tr}(U-U^\dagger)\cr
&&-a^2VW_6[{\rm Tr}\left(U+U^\dagger\right)]^2
     -a^2VW_7[{\rm Tr}\left(U-U^\dagger\right)]^2 \cr
&&-a^2 V W_8{\rm Tr}(U^2+{U^\dagger}^2) .
\ee
In addition to the chiral condensate, $\Sigma$, the action contains the 
low-energy Wilson constants $W_6$, $W_7$ and $W_8$ as unknown parameters 
\footnote{Note that we use the convention of \cite{DSV,ADSV} for the 
low energy constants $W_6$, $W_7$ and $W_8$. In the third entry of 
\cite{Sharpe} these constants are denoted by $-{W_6}'$, $-{W_7}'$ and 
$-{W_8}'$ respectively.}.
In the continuum the decomposition (\ref{decomb}) into a sum over
sectors of index $\nu$ is unambiguous \cite{LS}.
At nonzero lattice spacing the terms of order $a^2$ in the 
chiral Lagrangian, however, allow one to make different
decompositions corresponding to different definitions of $\theta$ and
$\nu$ at $a\neq0$.  As was shown in Ref.~\cite{ADSV}, the
specific decomposition (\ref{decomb}-\ref{lfull}) corresponds
exactly to the definition of $\nu$ given in Eq.~(\ref{defIndex}). 
Moreover, in the microscopic limit the real modes are predicted to be
in well seprated regions each width width of order $a^2 W_i/\Sigma\sim 1/V$.  
The gauge field configurations which contribute to the individual terms 
in the sum (\ref{decomb}) can thus be identified on the basis of the spectral
flow \cite{Heller0}. This provides a firm foundation for choosing this
particular lattice definition of the index.

To simplify our notation, we absorb the factor of $V
W_i$ into $a_i^2$ and the factor $V\Sigma $ into $m$, $z_k$ and $\lambda^5$ 
\begin{eqnarray}
  a^2 V W_i \to \ha_i^2 \qquad &&   mV \Sigma \to \hm  \cr
  z V \Sigma \to \hz\quad \, \quad  && \lambda^5 V\Sigma \to \hl^5.
\ee
The explicit expression for the quenched microscopic spectral density
can be derived following the procedure described in Ref.~\cite{DOTV}, 
now extended to the case of a source $z$ for the pseduoscalar
 density. In addition, much like at nonzero chemical potential 
\cite{SVfact,SVbos}, one must carefully take into account the 
convergence of the noncompact integrals. The resulting analytical formula was 
obtained in \cite{DSV,ADSV}: 
\be
\label{rho5}
\rho^\nu_5(\hl^5,\hm;\ha_i) & = & \frac{1}{\pi}{\rm Im} \ G^\nu(-\hl^5,\hm;\ha_i),
\ee
where 
\be
&&G^\nu(\hz,\hm;\ha_i) = \int_{-\infty}^\infty ds\int_{-\pi}^\pi
\frac{d\theta}{2\pi}
i\cos(\theta)e^{(i\theta-s)\nu} \cr
&& \hspace{-0.5cm}\times\!\exp[\!-\hm\sin(\theta)\!-\!i\! \hm\!\sinh(s)\!+\!i\hz\cos(\theta)
  \!-\!i\!(\hz\!-\!i\epsilon)\!\cosh(s) \nn \\
&&
+4\ha_6^2(-i\sin(\theta)+\sinh(s))^2
+4\ha_7^2(\cos(\theta)-\cosh(s))^2\cr
&&+2\ha_8^2(\cos(2\theta)-\cosh(2s))] 
 \nn \\
&& \hspace{-0.5cm}\times
\big(-\frac{\hm}{2}\sin(\theta)+
i\frac{\hm}{2}\sinh(s)+i\frac{\hz}{2}\cos(\theta)+i\frac{\hz}{2}\cosh(s)
\nn\\ 
&&\hspace{-0.2cm}
-4(\ha_6^2+\ha_7^2)(\sin^2(\theta)+\sinh^2(s))
\cr &&+2\ha_8^2(\cos(2\theta)+\cosh(2s)
+e^{i\theta+s}+e^{-i\theta-s})\!+\!\frac{1}{2}\big)  .
\label{res-result}
\ee
Existence of the integrals leads to the constraint $W_8>0$ 
(for $W_6=W_7=0$) as described in detail in Refs.~\cite{DSV,ADSV}. 
The sign of $W_8$ is also crucial for understanding if one approaches the
so-called Aoki phase of Wilson fermions \cite{Aokiclassic}. A positive
sign of $W_8$ (and $W_6=W_7=0$) indicates that the Aoki phase will
appear as the gap in the spectrum of $D_5$ closes. The analytic
predictions of WCPT for the spectral
density $D_5$ allow us to follow in detail how the gap disappears
as $m$ is taken to zero at fixed lattice spacing $a$.  

\section{Wilson chiral random matrix theory}
\label{sec:WRMT}

As we will make explicit use of it below, we give also here the
corresponding expressions in the equivalent Wilson chiral random matrix theory 
formulation \cite{DSV,ADSV}. The idea is to introduce the most general 
large-$N$ matrix that has conjugation properties similar to $\gamma_5$-Hermiticity.
Let this matrix be
\be
\tilde{D}_W= \mat aA & iW \\ iW^\dagger & aB \emat,
\label{w-diracOP}
\ee
where  
\be
A=A^\dagger \quad {\rm and}  \quad B^\dagger = B
\ee
are $(n+\nu) \times (n+\nu)$ and $n \times n$ complex matrices, 
respectively, and $W$ a rectangular complex matrix of
size $(n+\nu) \times n$. As in \cite{ADSV} we use tildes
to indicate quantities in Wilson chiral random matrix theory which
are analogues of those in the field theory. We consider the limit
$N = 2n + \nu \to \infty$ with $\tilde{m}N$, $\tilde{z}N$ and
$\tilde{a}^2N$ fixed. 

The matrix $\tilde{D}_W$ is $\tilde{\gamma}_5$-Hermitian with respect to
\be
\tilde{\gamma}_5 = {\rm diag}(1,\cdots, 1,-1, \cdots, -1) ~,
\ee
where the first $(n+\nu)$ entries are +1, and the remaining $n$
entries are -1.

Results based on the Wilson chiral random matrix theory are expected
to be universal, as in the case of the usual chiral random
matrix theory \cite{ADMN}. For simplicity one can therefore 
take the matrix elements to be distributed with Gaussian weight
\be
\label{P}
P(A,B,W) \equiv e^{-\frac {N}{2}{\rm Tr}[A^2+B^2] 
-N {\rm Tr} [ W W^\dagger]} .
\ee
The partition function of the Wilson chiral
random matrix theory is thus defined by
\be
\label{ZWRMT}
\tilde{Z}^\nu_{N_f} = \int\!\! dAdBdW \ \!\!\!\prod_{f=1}^{N_f}
\det(\tilde{D}_W+\tm_f+\tz_f\tilde{\gamma}_5) \ P . 
\ee
The matrix integrals are over the complex Haar measure.

In the limit $N \to \infty$ the partition function (\ref{ZWRMT}) coincides 
with that of (\ref{Zfull}) when $W_6=W_7=0$. It is straightforward to extend
the Wilson chiral random matrix theory so that also double-trace
terms corresponding to $W_6$ and $W_7$ are included \cite{ADSV,JO}. The
correspondence between parameters of Wilson chiral random matrix theory and
Wilson chiral perturbation theory is
\be
N\tm = \frac{m\Sigma V}{2}, \quad N\tz =
\frac{z\Sigma V}{2}, 
\quad \frac{N\ta^2}{2} = a^2 W_8 V .
\label{match}
\ee
Once this identification is made, one can equally well use the Wilson chiral random
matrix theory to derive spectral distributions. This has been demonstrated
explicitly in \cite{AN,KVZ}. Moreover, since no compact expressions
are presently known for individual eigenvalue distributions, it is straightforward
to generate such eigenvalue distributions numerically using the Gaussian integrals
of Wilson chiral random matrix theory. This will be used below.

\section{The lattice setup}
\label{sec:lat}

Here we give a few details of the lattice simulations. For this initial
study we work in the quenched approximation. It is advantageous
to use a pure gauge action which gives configurations with a fairly unique
topological charge, {\it i.e.,} a pure gauge action that suppresses
so-called dislocations. Experience from simulations with domain wall
fermions, for which the dislocations tend to increase the residual chiral
symmetry breaking at finite fifth dimensional extent $L_s$, suggests that
the Iwasaki gauge action \cite{Iwasaki} would be a good choice.  This
action is fairly well studied and appears to have a rather smooth continuum
limit. Our lattice parameters were chosen using $r_0/a$ values from
interpolation formulae given in \cite{Iwasaki_r0.a,Iwasaki_r0.b} and using
$r_0 = 0.5$ fm to set the physical scale.

We used a mixture of heatbath and overrelaxation updates with, as advocated
in Ref.~\cite{Iwasaki_r0.b}, $N_{or} \approx 1.5 r_0/a$ overrelaxation
sweeps for each heatbath sweep.  We saved configurations separated by 100
heatbath sweeps for our measurements. We found no detectable
autocorrelations in the ensembles generated. Experience indicates that
for studies of the $\epsilon$-regime of chiral perturbation theory,
physical lattice sizes of $L = 1.5$ to 2 fm are most useful. The lattice
size is large enough so that several Dirac eigenvalues are governed by
the $\epsilon$-regime of CPT, but it is small enough that sufficient
statistics in sectors of given, small topological charge can be
obtained. Some relevant parameters of the three ensemble considered in
this study are summarized in Table~\ref{tab:configs}.

\begin{table}
\begin{tabular}{|c|l|c|c|c|c|c|}
\hline
Ensemble & $\beta_{Iw}$ & size & $r_0/a$ & $a$ [fm] & $L$ [fm] & \# cfgs \\
\hline
A & 2.635 & $16^4$ & 5.37 & 0.093 & 1.5 & 6500 \\
B & 2.635 & $20^4$ & 5.37 & 0.093 & 1.9 & 3000 \\
C & 2.79  & $20^4$ & 6.70 & 0.075 & 1.5 & 6000 \\
\hline
\end{tabular}
\caption{Parameters of the pure gauge configurations considered.
They are generated with the Iwasaki gauge action.}
\label{tab:configs}
\end{table}

\subsection{Spectral flow}

In order for the match to Wilson chiral perturbation theory to be 
valid the assignment of index $\nu$ to a given gauge field
configuration must be obtained from Eq.~(\ref{defIndex}) using 
either {\sl a)} the (small) real eigenvalues of the Wilson-Dirac 
operator, together with the chiralities of the corresponding 
eigenvectors or equivalently {\sl b)} by counting the number of net
crossings of zero of the eigenvalues of the Hermitian Wilson-Dirac
operator as a function of the mass parameter $m_0$ \cite{Heller0} and 
weighting their contribution by the sign of the slope at the crossing.
 Since we did not have an implementation of the Arnoldi
algorithm to compute the (complex) eigenvalues of the Wilson-Dirac
operator, including the relevant real eigenvalues, we used the spectral
flow strategy.

The index determined from the real eigenvalues of the Wilson-Dirac operator, 
or equivalently from the spectral flow, can depend on the range of the real
eigenvalues considered, or equivalently, the value of $-m_{cut}$ at which
the spectral flow is terminated. With $-m_c$ being the critical mass, at
which the pion formed from Wilson quarks becomes massless, $m_{cut}$ should
be sufficiently larger than $m_c$, but smaller than the value at which the
doubler fermions become light. For our computations we use one HYP smearing
of the gauge fields \cite{HYP} in the construction of the Wilson-Dirac operator to
help further suppress lattice artifacts.  Then the choice $m_{cut} = 1.0$
is safely away from both $m_c$ and the doubler region. With the Iwasaki
gauge action and its suppression of dislocations, crossings in the
neighborhood of $m_{cut}$ are rare, but they do occur occasionally. Hence a
change in $m_{cut}$ would change the index on a few configurations.

\begin{table}
\begin{tabular}{|c|c|l|c|c|}
\hline
Ensemble & size & $\beta_{Iw}$ & $m_0$ & $\#$ configs  \\
 &      & & & $\nu=0,1,-1$ \\
\hline
A & $16^4$ & 2.635 & $-0.216$ & 1246, 1088, 1045 \\
B & $20^4$ & 2.635 & $-0.216$ & 379, 319, 322 \\
C & $20^4$ & 2.79 & $-0.178$ & 1172, 990, 988      \\
C & $20^4$ & 2.79 & $-0.184$ & 1172, 990, 988   \\ 
\hline
\end{tabular}
\caption{\label{tab:lat}}
\end{table}

\section{The Spectrum of $D_5$: lattice simulations}
\label{sec:results}

In this section we compare the microscopic spectral density of $D_5$
as predicted by WCPT for the quenched case to the lattice simulations
described above.

The values of $m$, $\lambda^5$ and $a$ given in the figures refer to
$\hm$, $\hl^5$ and $\ha_8$. For all plots except Figure
\ref{fig:a6a7dep} we have set $\ha_6=\ha_7=0$.

\subsection{Volume scaling}

Our first test of the analytic predictions for the microscopic
spectrum of $D_5$ concerns the finite volume scaling at fixed lattice spacing. 
That is, the scaling of $\hm$, $\hl^5$ and $\ha_8$ with $V$ for fixed $a$.
To this end we consider the two sets of data with $\beta_{Iw}=2.635$
and $m_0=-0.216$, see Table \ref{tab:lat}. Because of the larger
statistics obtained we first   
match the $16^4$ lattice data with $\nu=0$ to the 
prediction (\ref{res-result}). The result is displayed in the top
panel of Figure \ref{fig:vol-scalingnu0}. We observe that 
it is possible to make a reasonable fit to the
data. The range over which the predictions match the data
is limited to the lowest two eigenvalues, beyond which the increasing
slope of the data indicates that the $p$-regime effects set in. 
Given the best values found in the fit we then scale $\hm$ by
$20^4/16^4$ and $\ha_8$ by $20^2/16^2$ to obtain the analytic
prediction for the spectral density obtained on the $20^4$ lattice 
at the same value of the coupling and lattice mass. This parameter 
free prediction is plotted against the lattice data in the lower panel
of Figure \ref{fig:vol-scalingnu0}. We observe that this finite-volume
scaling of the analytic prediction is consistent with the $20^4$
data. The limited statistics does not allow us to resolve the detailed
structure of the analytic prediction.

\begin{center}
\begin{figure}[t!]
\includegraphics[width=8cm,angle=0]{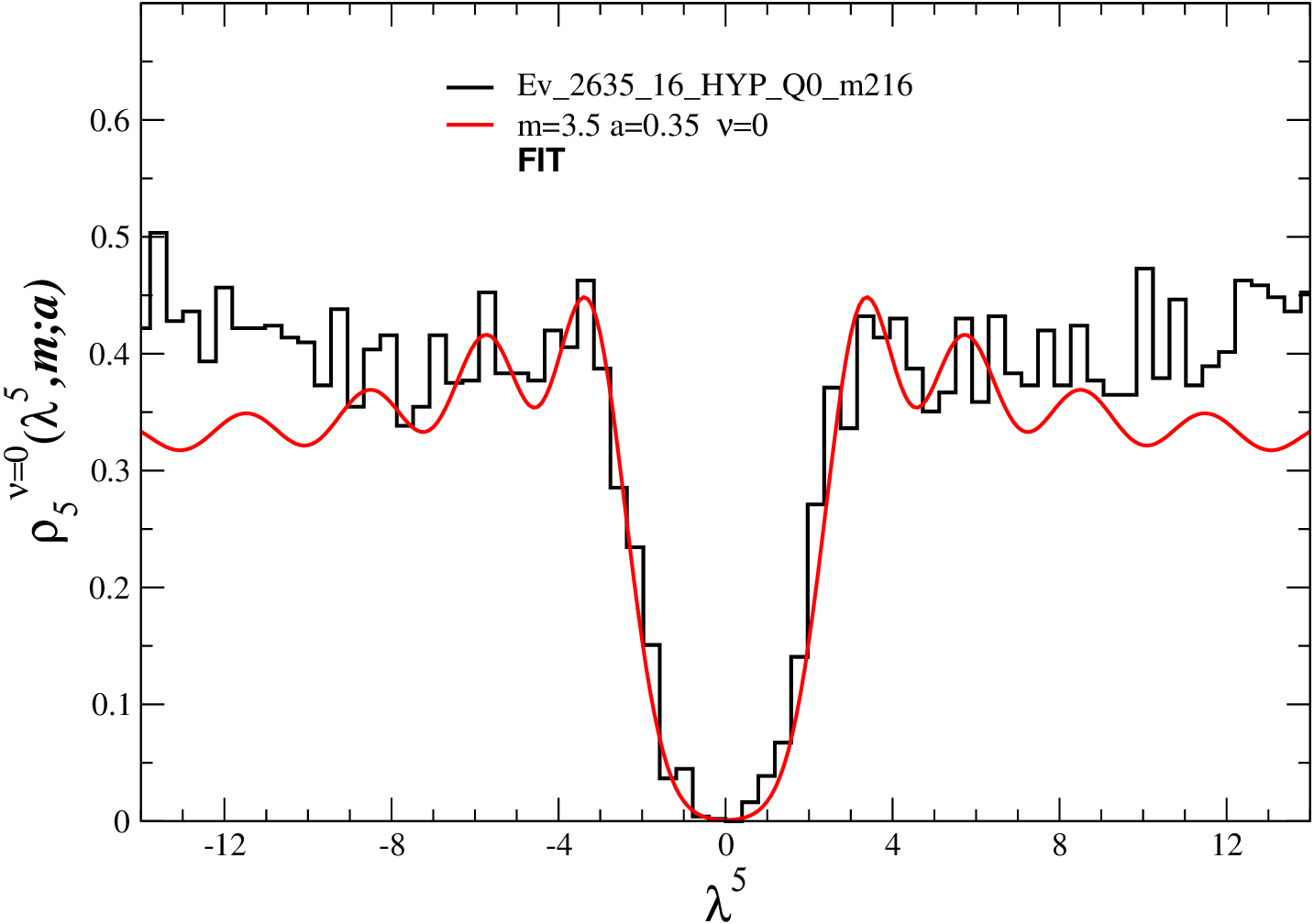}
\vspace{10mm}
\vfill
\includegraphics[width=8cm,angle=0]{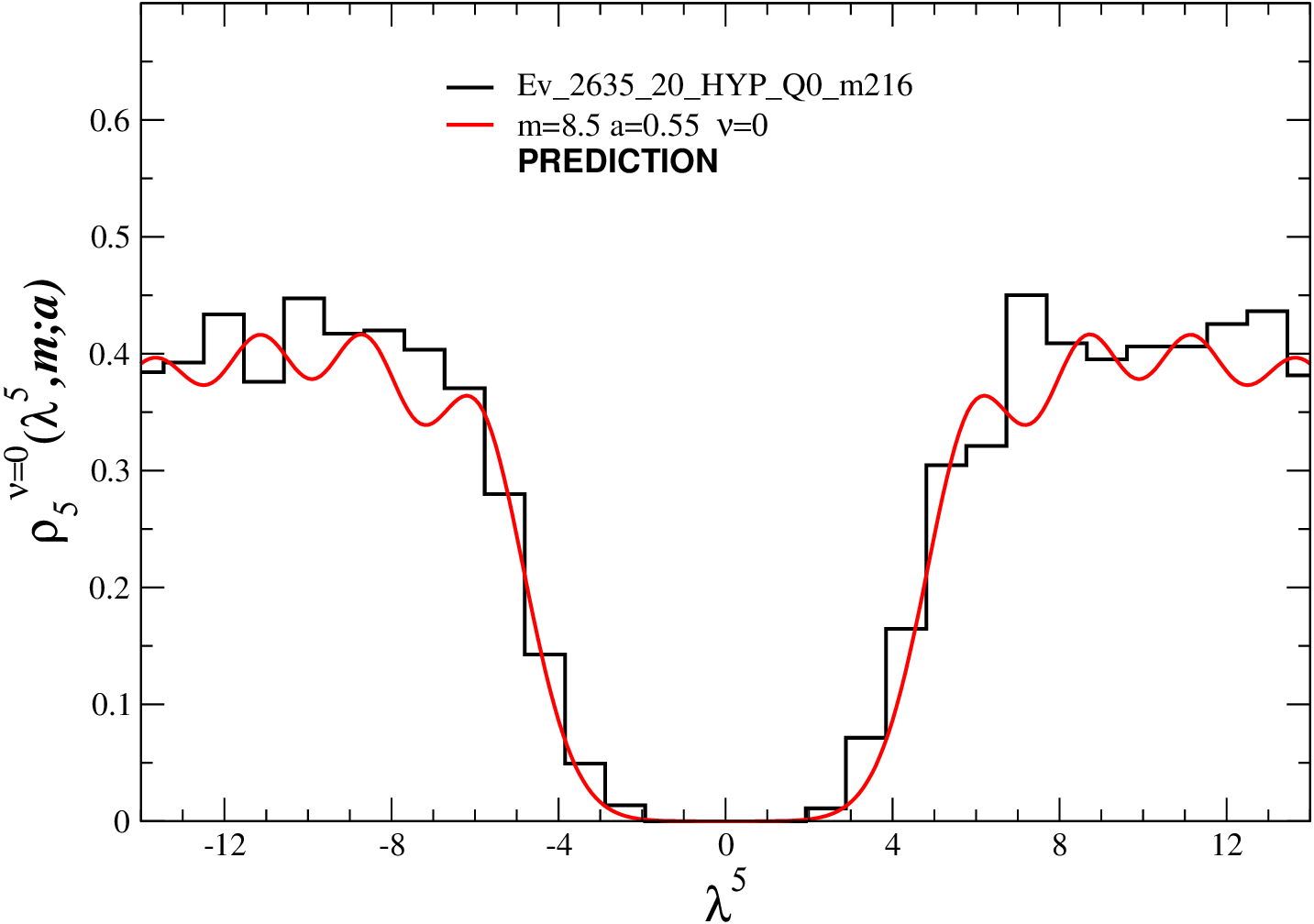}
\caption{\label{fig:vol-scalingnu0} The eigenvalue density of $D_5$ in the
  sector with $\nu=0$ for
  $V=16^4$ ({\bf top}) and $V=20^4$ ({\bf bottom}) both with
  $\beta_{Iw}=2.635$ and $m_0=-0.216$. The $x$-axis has been rescaled by
  $\Sigma V$, where $\Sigma 16^4=157.5$. In the top plot the smooth 
  red curve displays a fit of the analytic prediction for the microscopic
  density of $D_5$ to the $16^4$ data. The fit values are $\hat{m}=3.5$
  and $\hat{a}_8=0.35$ ($\hat{a}_6=\hat{a}_7=0$). In the plot below
  the smooth red line shows the prediction for the $20^4$ data
  ($\hat{m}=8.5$ and $\hat{a}_8=0.55$) obtained from finite volume
  scaling of the $16^4$ values. We observe that the
  finite volume scaling at fixed lattice spacing is well respected.}
\end{figure}
\end{center}

\begin{center}
\begin{figure}[t!]
\includegraphics[width=8cm,angle=0]{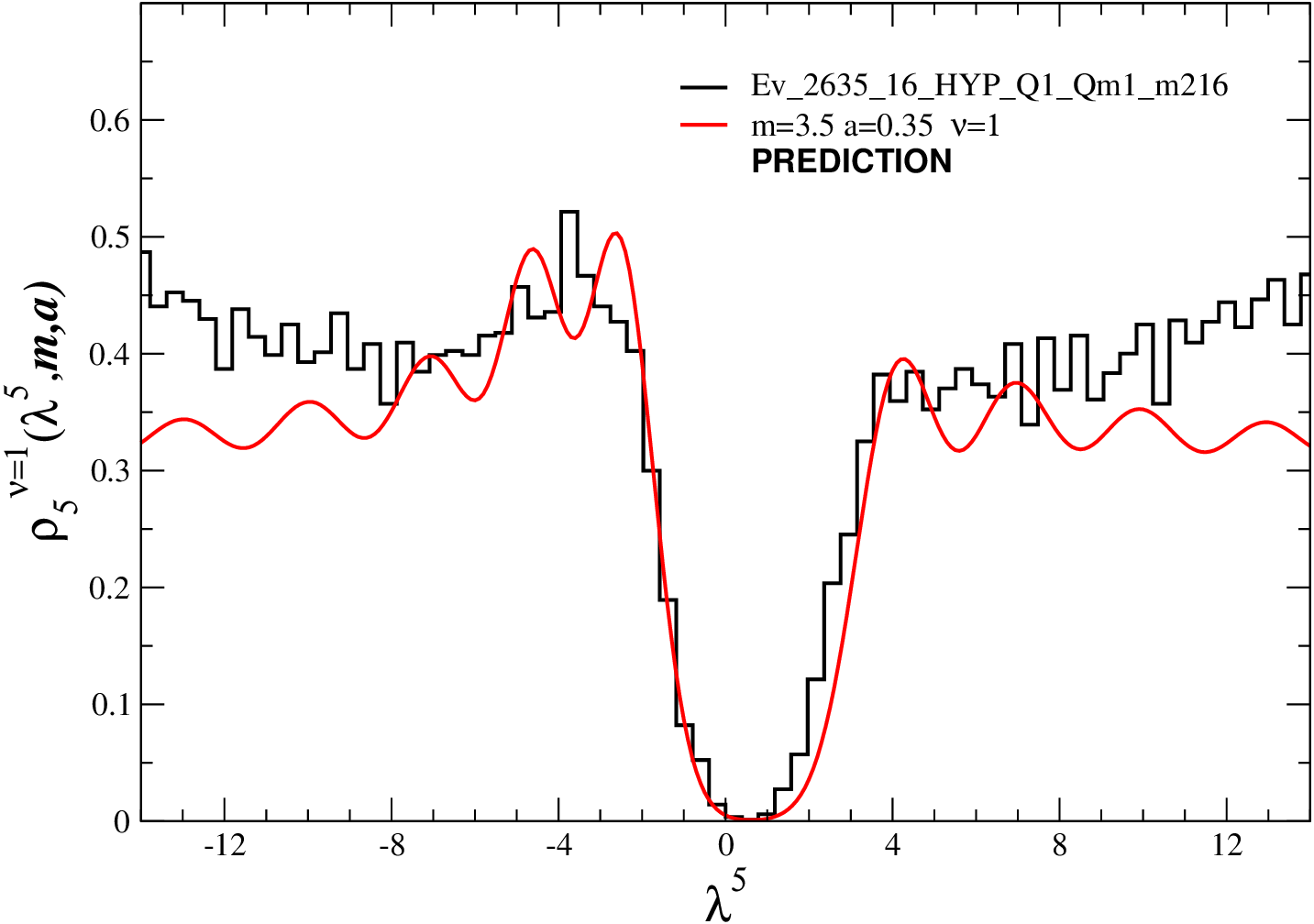}
\vspace{10mm}
\vfill
\includegraphics[width=8cm,angle=0]{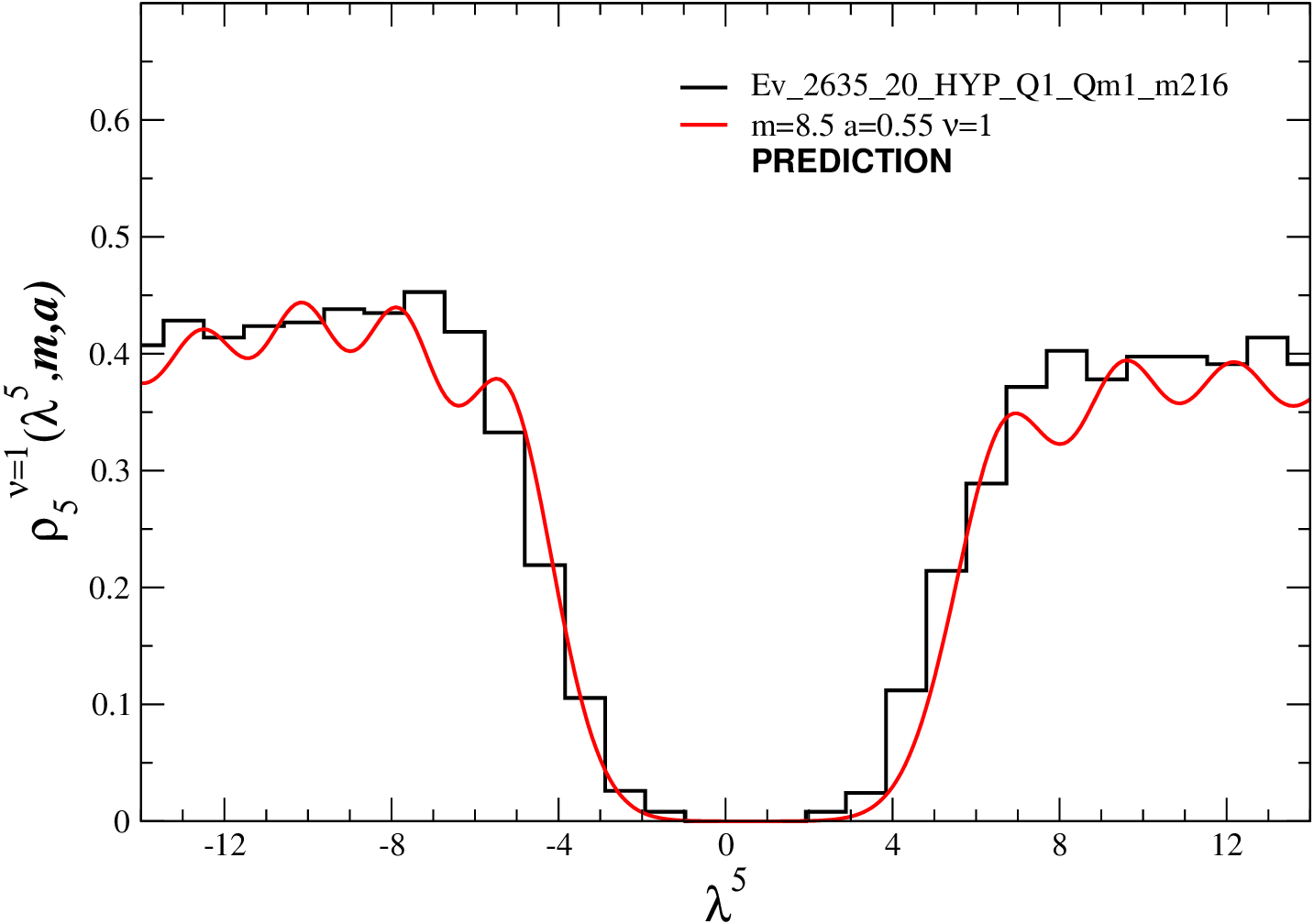}
\caption{\label{fig:vol-scalingnu1} The eigenvalue density of $D_5$ in the
  sector with $\nu=1$ for
  $V=16^4$ ({\bf top}) and $V=20^4$ ({\bf bottom}) both with
  $\beta_{Iw}=2.635$ and $m_0=-0.216$. The $x$-axis has been rescaled by
  $\Sigma V$, where $\Sigma 16^4=157.5$. The smooth 
  red curves displays the analytic prediction for the microscopic
  density of $D_5$. The values, $\hat{m}=3.5$ and $\hat{a}_8=0.35$ for
  $16^4$ and $\hat{m}=8.5$ and $\hat{a}_8=0.55$ are obtained from the
  fit in the top panel of Figure \ref{fig:vol-scalingnu0} and by finite volume
  scaling from $16^4$ to $20^4$ ($\hat{a}_6=\hat{a}_7=0$).}
\end{figure}
\end{center}

\vspace{-13mm}
Likewise for $|\nu|=1$ we can test the analytic prediction with the
parameters fixed from the fit to the $16^4$ and $\nu=0$ data. The 
result is displayed in Figure \ref{fig:vol-scalingnu1}. Note that the
analytic prediction for $-\nu$ is obtained from the one for $\nu$ by
flipping the sign of $\hl^5$. To make the most of our statistics we
therefore overlay the data obtained for $\nu=1$ and $\nu=-1$ flipping
the sign of the eigenvalues of the latter. This is done throughout
this paper. We observe that the general structure of the 
$16^4$ $\nu=1$ data is reproduced by the analytic prediction. 
Also the overall features of the $20^4$ $\nu=1$ data are followed 
by the parameter free analytic prediction which in this case follows 
from finite volume scaling.

We conclude that $W_6=W_7=0$ and $W_8>0$ is consistent with the
lattice data at $\beta_{Iw}=2.635$ and that the volume scaling 
at fixed lattice spacing predicted by Wilson chiral perturbation 
works well.

\subsection{Mass scaling}

Our second test of the analytic predictions for the spectrum of $D_5$ concerns 
the scaling with the quark mass at fixed lattice spacing. For this we choose a 
$20^4$ lattice with $\beta_{Iw}=2.79$ and two masses $m_0=-0.178$ respectively 
$m_0=-0.184$.

We first consider the $m_0=-0.184$ data with $\nu=0$. We make a fit and find 
the best values of $\hm=3.5$, $\Sigma V=216$ and $\ha_8=0.35$. (The fact that 
$\hat{m}$ and $\hat{a}$ takes the same value as for the $16^4$ lattice at 
$\beta_{Iw}=2.635$ is accidental.) For these values the analytic curve matches 
the data well over the lowest four eigenvalues, see the top panel of 
Figure \ref{fig:fit2smallmass}.   

Given these values of $\hm$ and $\ha_8$ we
then test the predicted density for $\nu=1$ against the data. As can 
be observed from the lower panel of Figure \ref{fig:fit2smallmass}
this again works well for the lowest four eigenvalues. Finally, we
consider also the higher mass data at the same value of the
coupling. Since the lattice mass has changed from $m_0=-0.184$ to
$m_0=-0.178$ we have $\delta m = 0.006$ so that
$\delta m \Sigma V = 0.006*216 = 1.3$. Hence, we predict that the
$\beta_{Iw}=2.79$ data with $m_0=-0.178$ should be matched with
$\hm=4.8$ and $\ha=0.35$. As can be seen from Figure
\ref{fig:fit2largemass} this parameter free prediction works well 
for the $\nu=0$ as well as for the $|\nu|=1$ sector. Figure 
\ref{fig:Q2} compares the corresponding parameter free prediction for 
$|\nu|=2$ against the data.

We conclude that $W_6=W_7=0$ and $W_8>0$ is consistent with the
$20^4$ lattice data with $\beta_{Iw}=2.79$ and that the mass scaling 
at fixed lattice spacing implied by WCPT works 
well. For the influence of $W_6$ and $W_7$ see below.

\begin{center}
\begin{figure}[t!]
\includegraphics[width=8cm,angle=0]{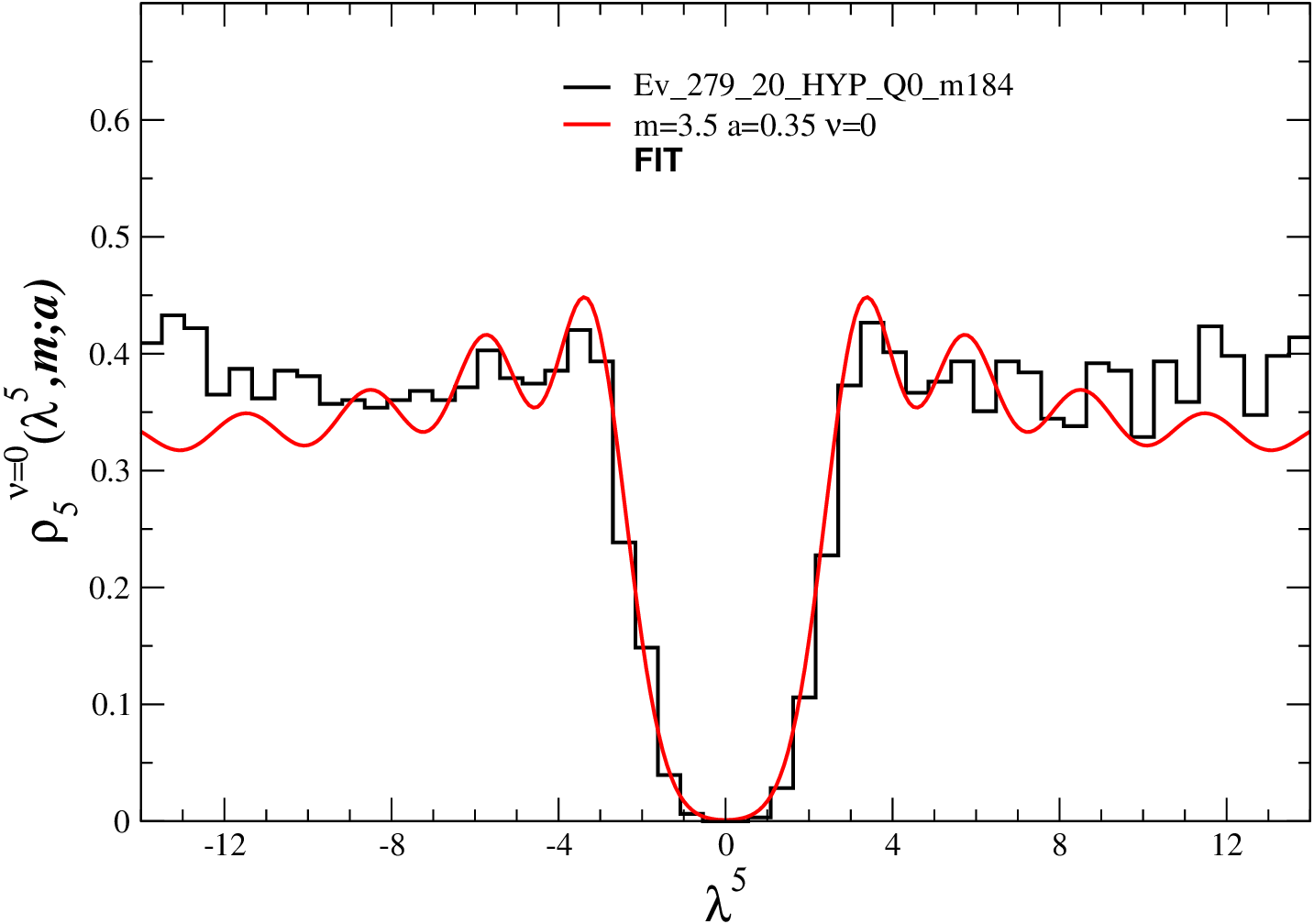}
\vspace{10mm}
\vfill
\includegraphics[width=8cm,angle=0]{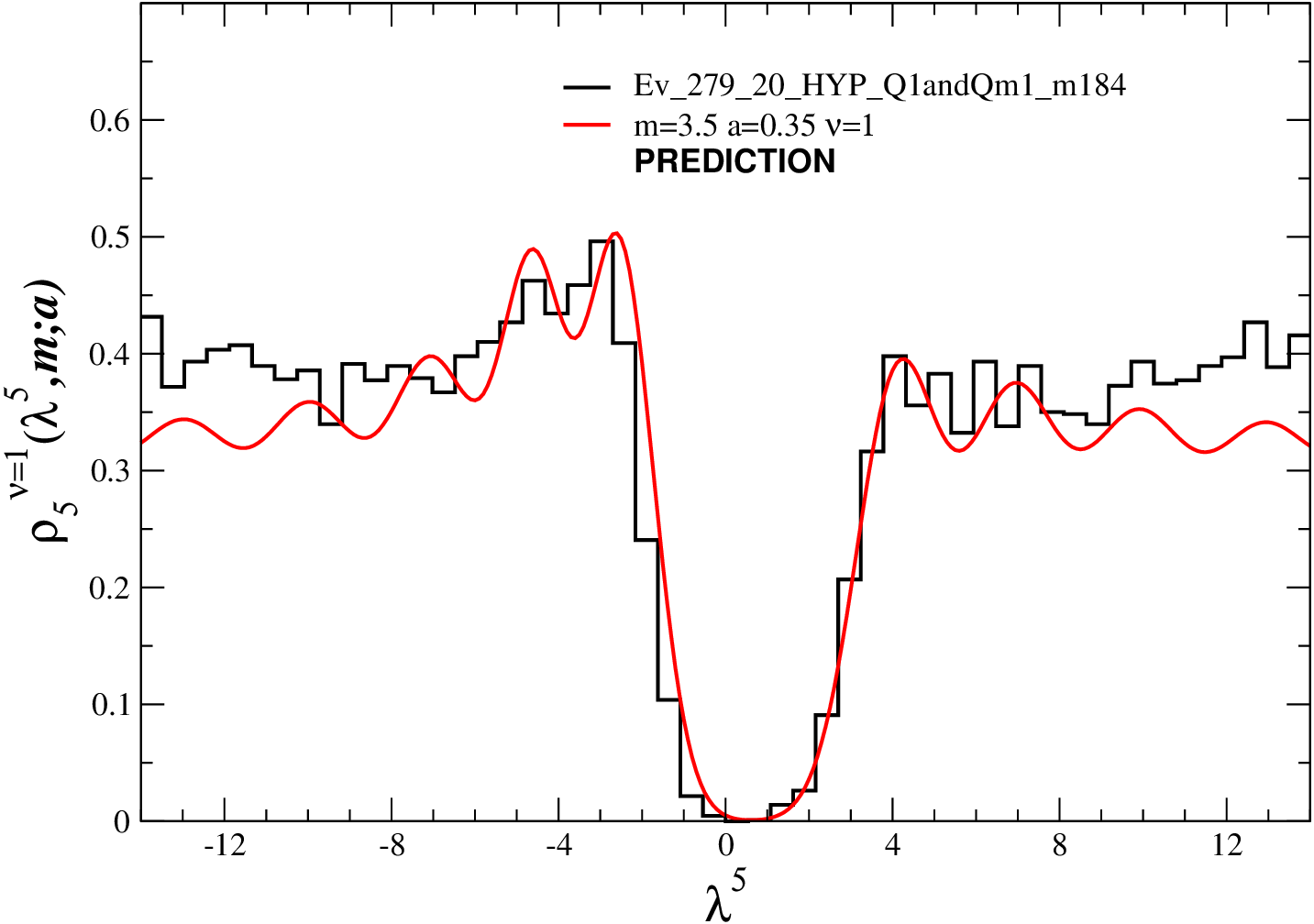}
\caption{\label{fig:fit2smallmass} 
The eigenvalue density of $D_5$ for $\beta_{Iw}=2.79$, $V=20^4$ and $m_0=-0.184$ {\bf top:}
$\nu=0$ and {\bf bottom:} $|\nu|=1$. For the $\nu=0$ data we find the
best values $\hm=3.5$, $\ha_8=0.35$ and $\Sigma V =216$. With these
values we then test the prediction against the $|\nu|=1$ data.}
\end{figure}
\end{center}

\begin{center}
\begin{figure}[t!]
\includegraphics[width=8cm,angle=0]{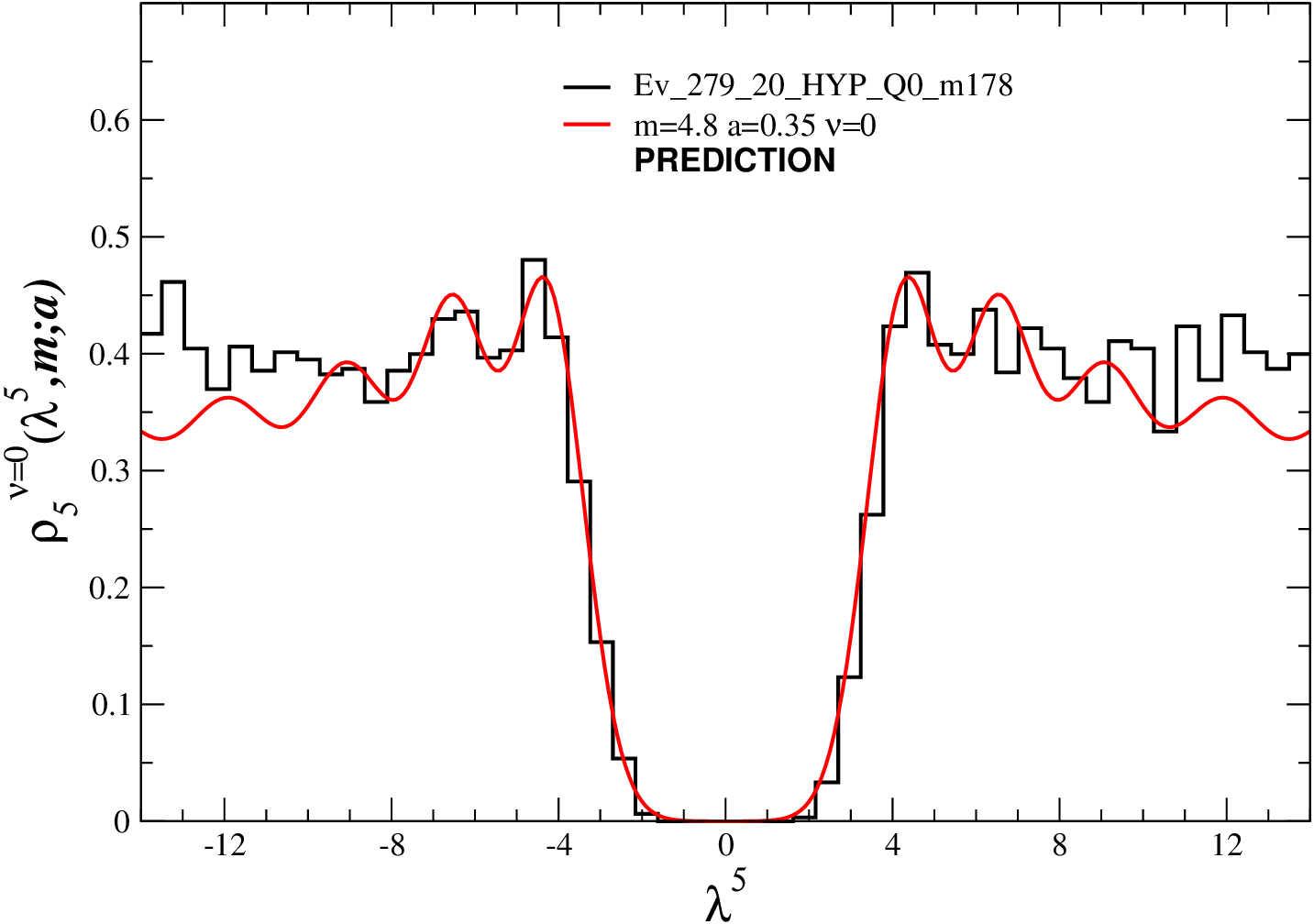}
\vspace{10mm}
 \vfill
\includegraphics[width=8cm,angle=0]{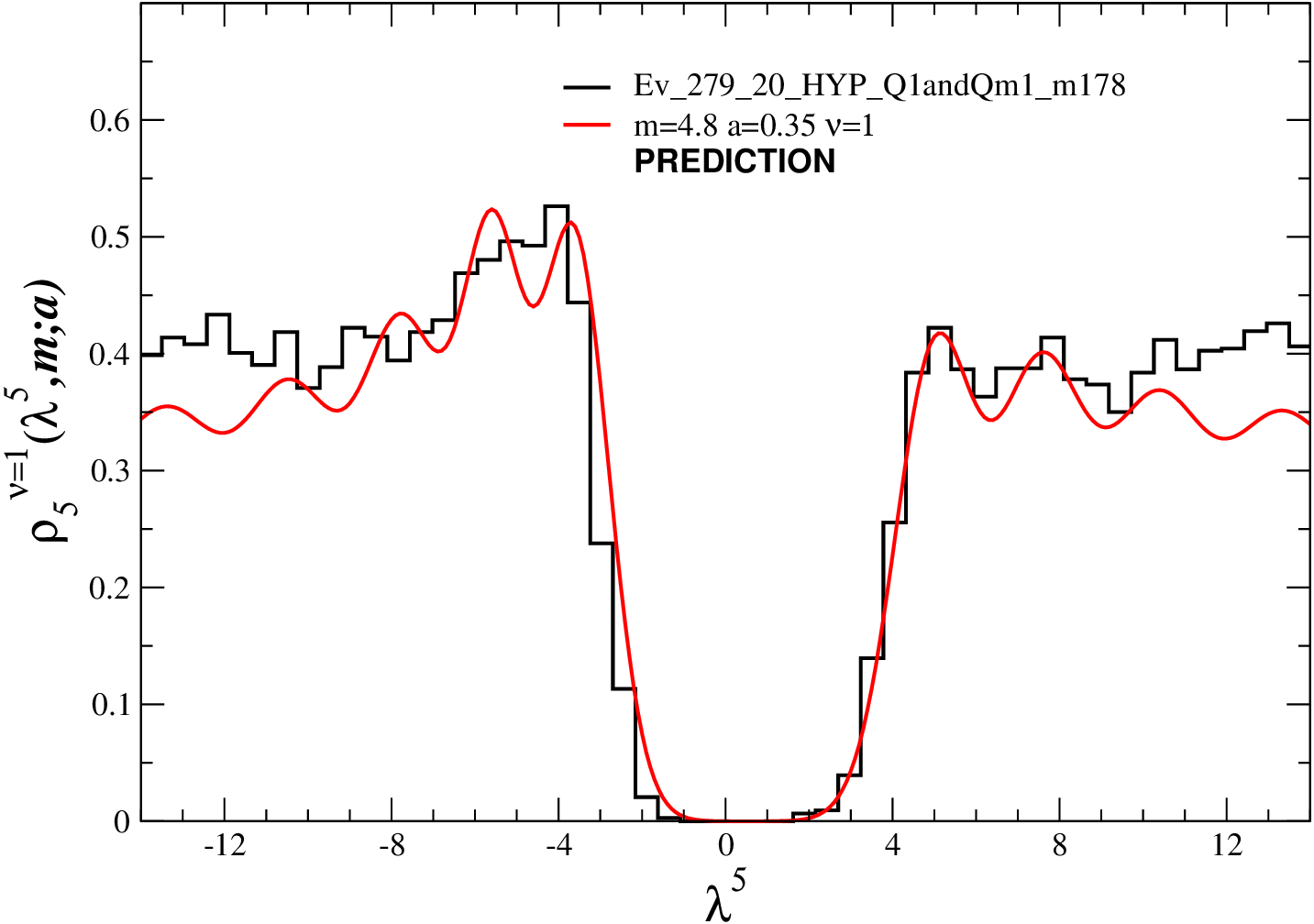}
\caption{\label{fig:fit2largemass} 
The eigenvalue density of $D_5$ for $\beta_{Iw}=2.79$, $V=20^4$ and $m_0=-0.178$ {\bf top:}
$\nu=0$ and {\bf bottom:} $|\nu|=1$. We use the value $\Sigma V=216$
(found for $m_0=-0.184$) and $\delta m=-0.006$ to determine the 
value $\hm=4.8$ for the $\beta_{Iw}=2.79$, $V=20^4$,
$m_0=-0.178$ data. Since the coupling is unchanged also $\ha_8=0.35$
is fixed by the previous fit. The parameter free predictions for
$\nu=0$ and $|\nu|=1$ are then plotted against the data. We observe that 
the mass scaling at fixed lattice spacing works well.}
\end{figure}
\end{center}

\begin{center}
\begin{figure}[t!]
\includegraphics[width=8cm,angle=0]{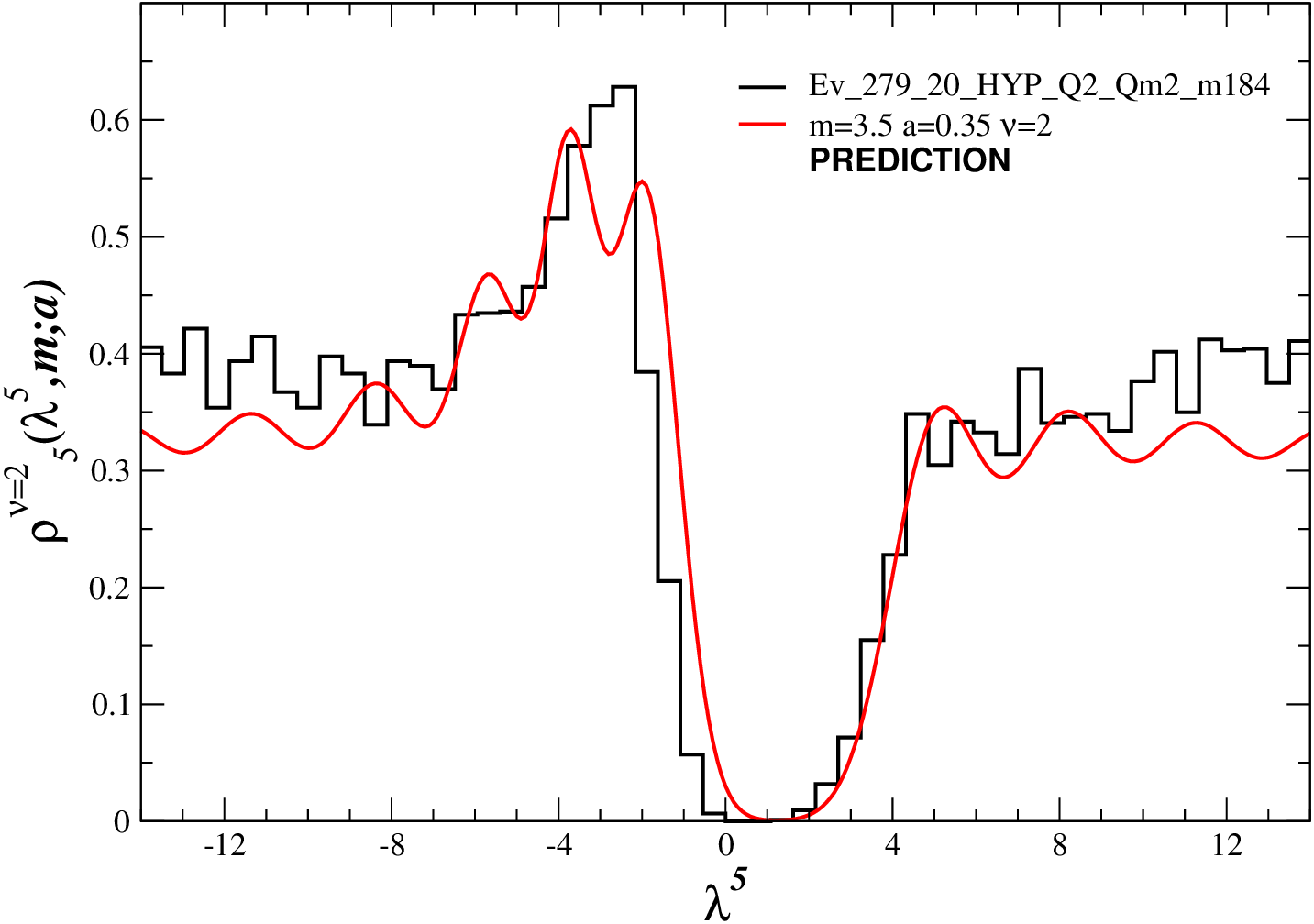}
\vspace{10mm}
 \vfill
\includegraphics[width=8cm,angle=0]{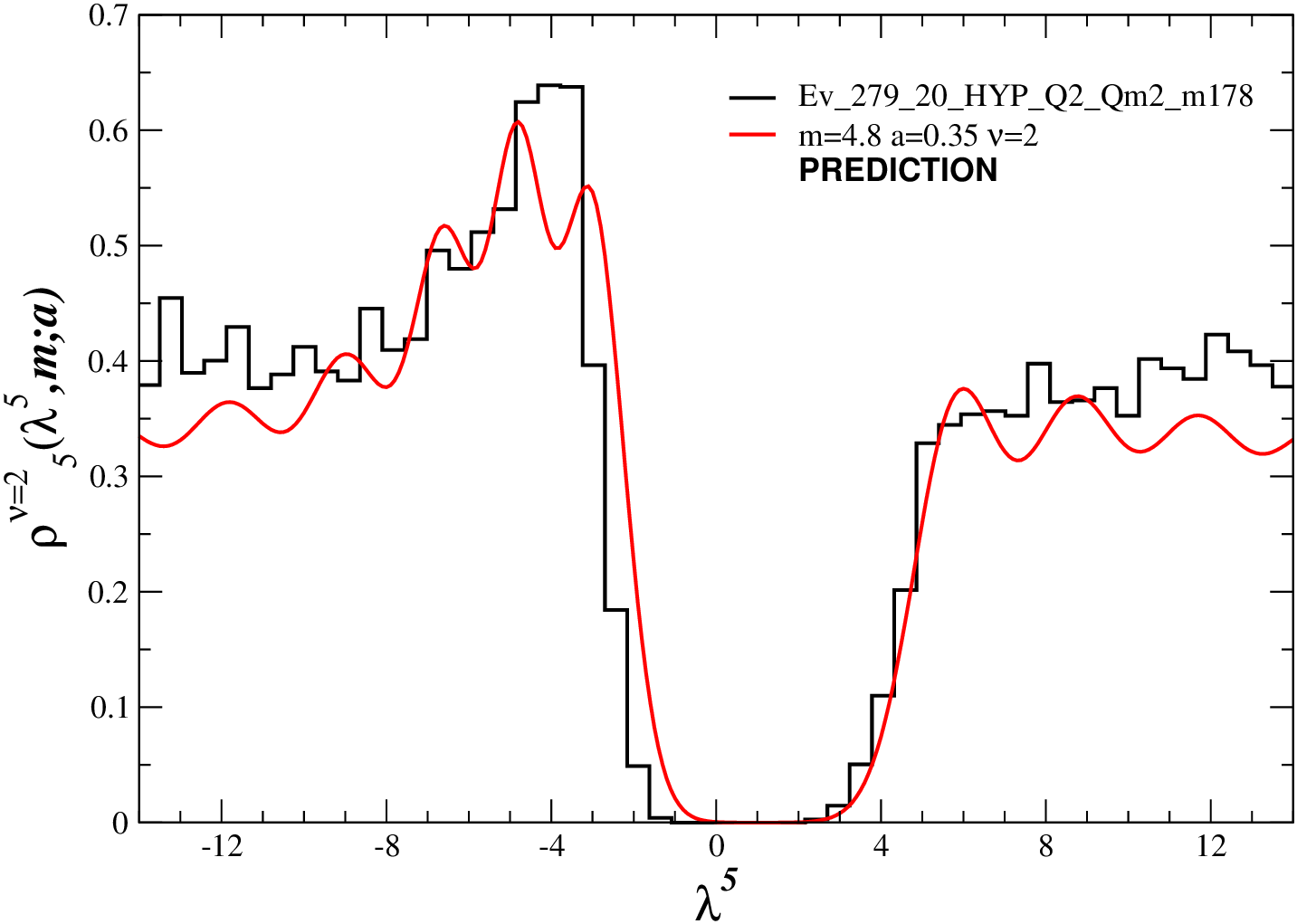}
\caption{\label{fig:Q2} 
The eigenvalue density of $D_5$ for $\beta_{Iw}=2.79$, $V=20^4$ and $|\nu|=2$  
{\bf top:} $m_0=-0.184$ and {\bf bottom:} $m_0=-0.178$. The smooth curves are the 
parameter free predictions from WCPT (all parameters are fixed from the fit to 
the $\nu=0$, $m_0=-0.184$ data).}
\end{figure}
\end{center}

\subsection{The $\hm$ and $\ha_8$ dependence}

Above we have obtained best fits of the analytic prediction,
Eq.~(\ref{res-result}), to the lattice data by varying $\hm$
and $\ha_8$. Here we demonstrate the strong sensitivity of the
analytic predictions for the microscopic eigenvalue density of $D_5$
with fixed index to the values of $\hm$ and 
$\ha_8$. Figure \ref{fig:mANDadep} give examples of this strong dependence 
which allow for a precise determination of the best values for the match 
to the data. Figure \ref{fig:a0} compares the best fit to the continuum, 
$a=0$, curve. In this way we estimate 
the error on $\ha_8$ in our data sets to be at most $0.05$ and the 
error on $\hm$ to be below $0.5$. With increased statistics
on slightly larger lattices at slightly smaller lattice spacing it 
should be possible to determine $\hm$ and $\ha_8$ with high
accuracy.

\begin{center}
\begin{figure}[t!]
\includegraphics[width=8cm,angle=0]{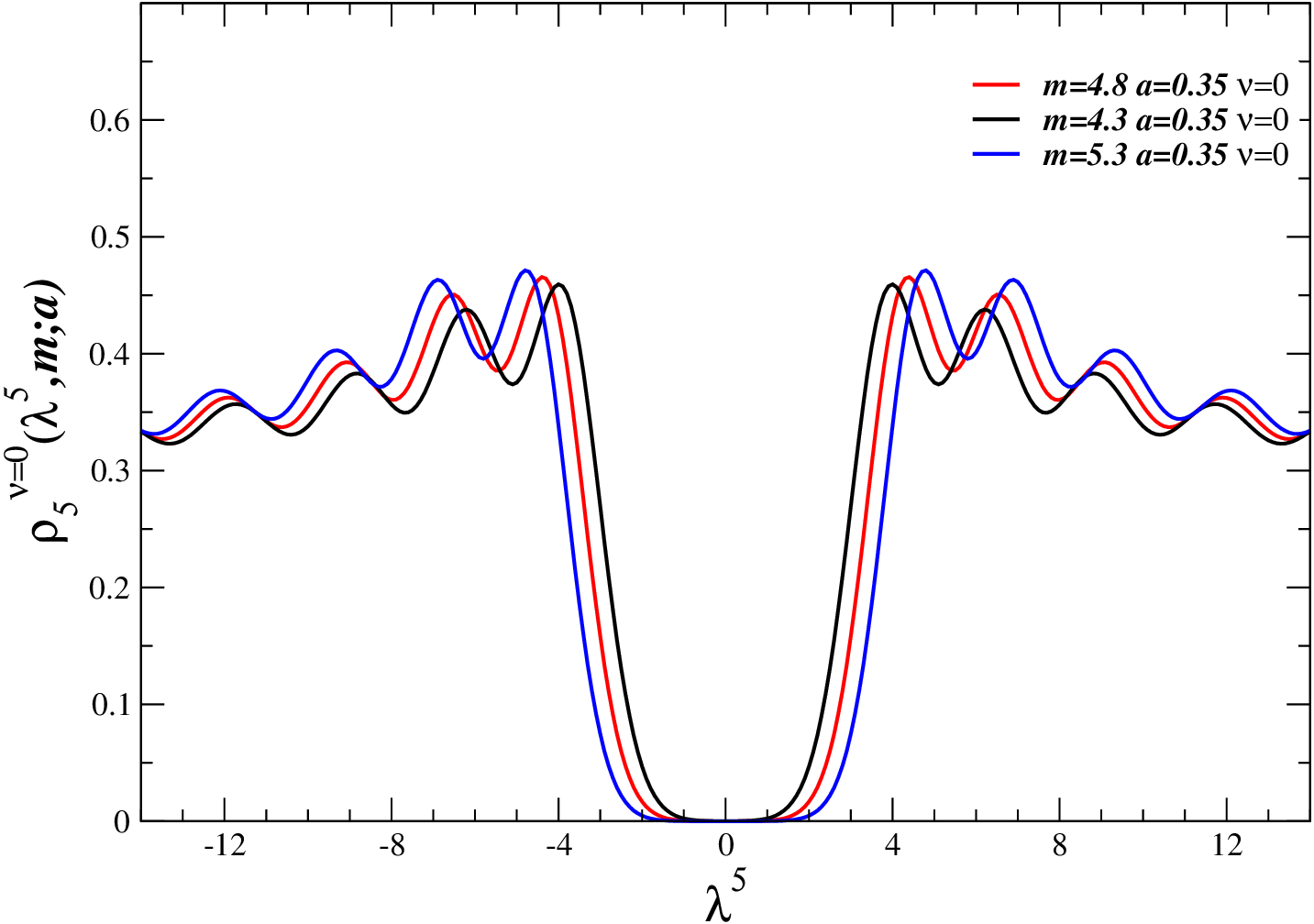}
\vspace{10mm}
 \vfill
\includegraphics[width=8cm,angle=0]{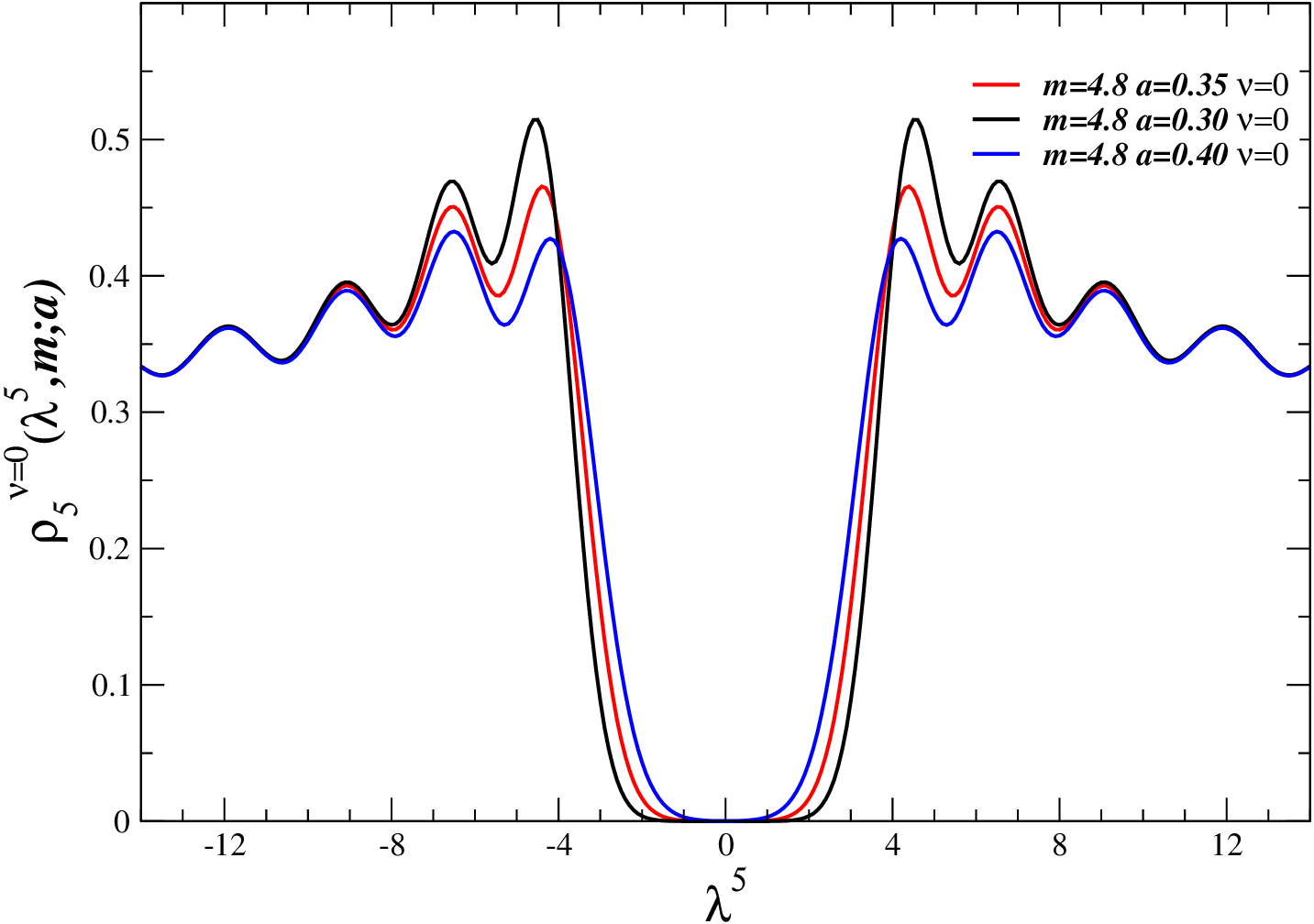}
\caption{\label{fig:mANDadep} 
The dependence of the eigenvalue density of $D_5$ for $\nu=0$ on {\bf top:} the
quark mass $\hm$ and on {\bf bottom:} the lattice artifacts $\ha_8$. The
central (Red) line in both plots correspond to the (Red) analytic
curve which is plotted 
against the data in the top panel of Figure \ref{fig:fit2largemass}.}
\end{figure}
\end{center}

\begin{center}
\begin{figure}[t!]
\includegraphics[width=8cm,angle=0]{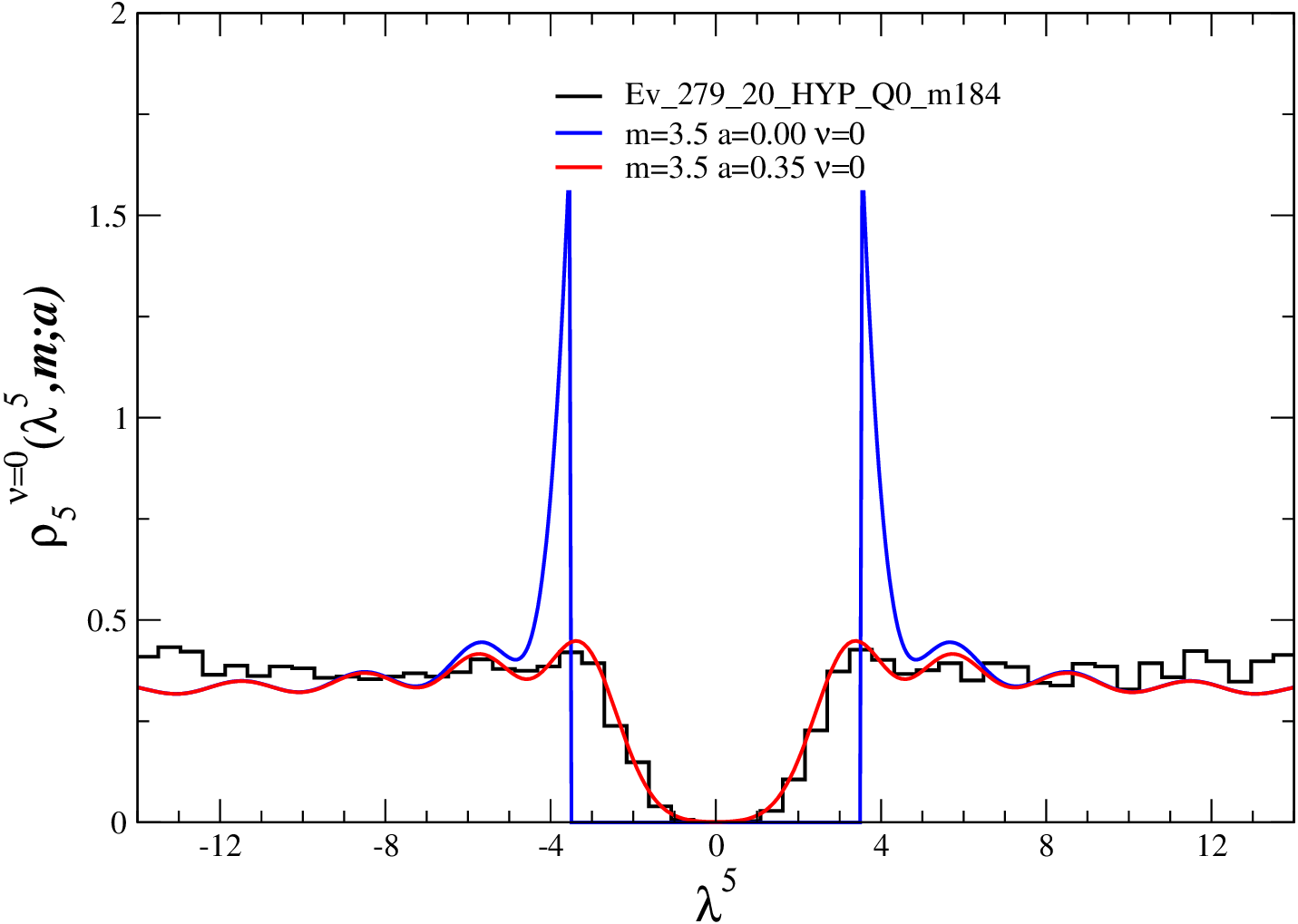}
\caption{\label{fig:a0} 
The eigenvalue density of $D_5$ for $\nu=0$. Shown are the lattice data and   
best fit, previously displayed in the top panel of Figure
 \ref{fig:fit2smallmass},  along with the $a=0$ result at the same value of 
$\hat{m}$. Clearly it is essential to  
include the order $a^2$ effects in WCPT in order to match the data. Note that 
scale on the $y$-axis is different from that in the other plots.}
\end{figure}
\end{center}

\subsection{Effects of $W_6$ and $W_7$}

The effect of $W_6<0$ or $W_7<0$ is to smear out the analytic predictions 
so that the bumps in the curves become less pronounced. 
In Figure \ref{fig:a6a7dep} we give an example of the effect of $\ha_6^2$ 
and $\ha_7^2$. The more oscillating curve in both plots corresponds to the 
curve in the top panel of Figure \ref{fig:fit2largemass} which has 
$\ha_6=\ha_7=0$. As a less pronounced structure of the oscillations 
due to the individual eigenvalues is unfavored by the lattice data 
we conclude that $|\ha_6^2|$ and $|\ha_7^2|$ appear to be smaller 
than $0.01$. Note that $\ha_i$ only enters as squares in the action 
(\ref{lfull}), that is why it is natural to give the bound on the square. 

\begin{center}
\begin{figure}[t!]
\includegraphics[width=8cm,angle=0]{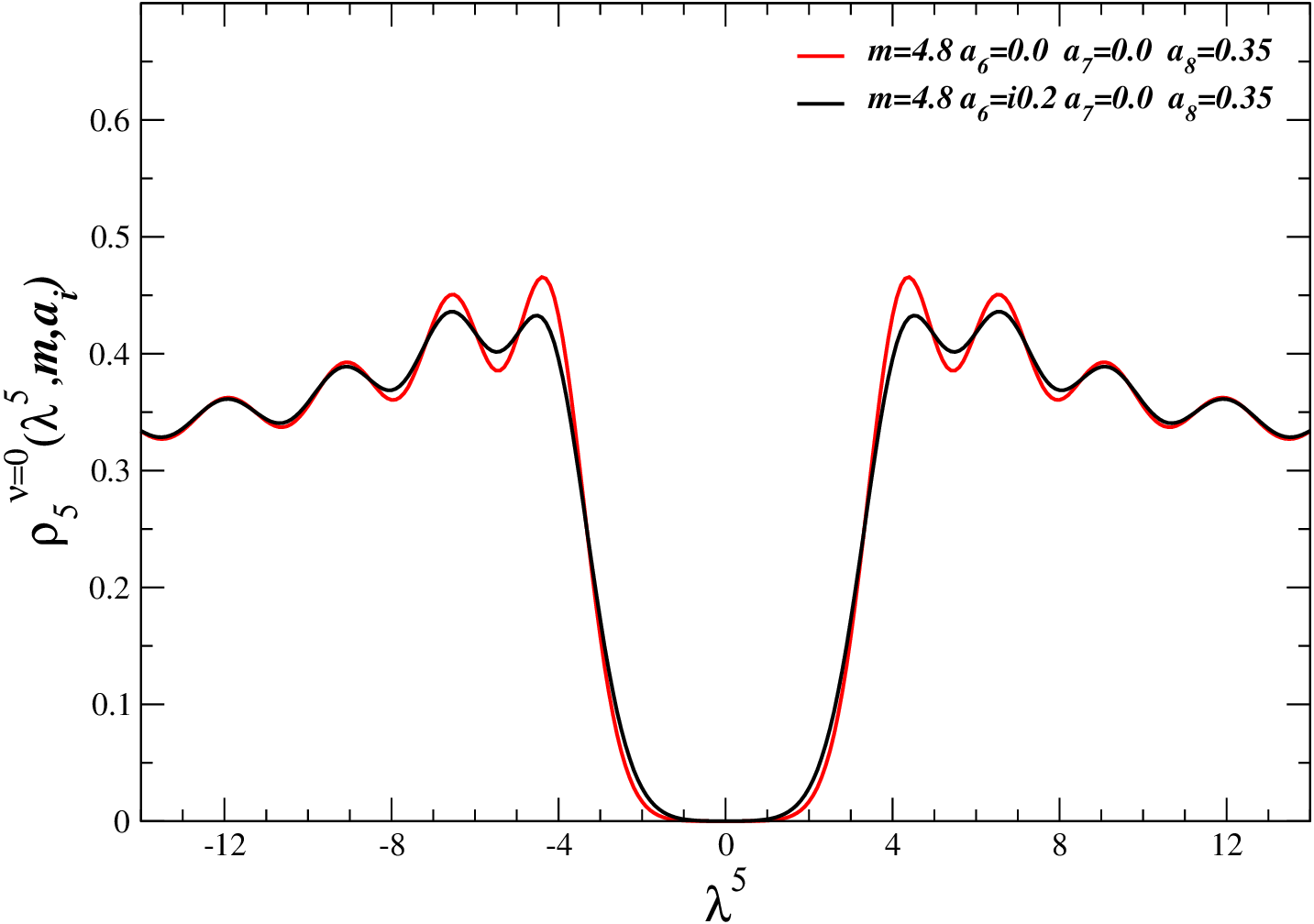}
\vspace{10mm}
 \vfill
\includegraphics[width=8cm,angle=0]{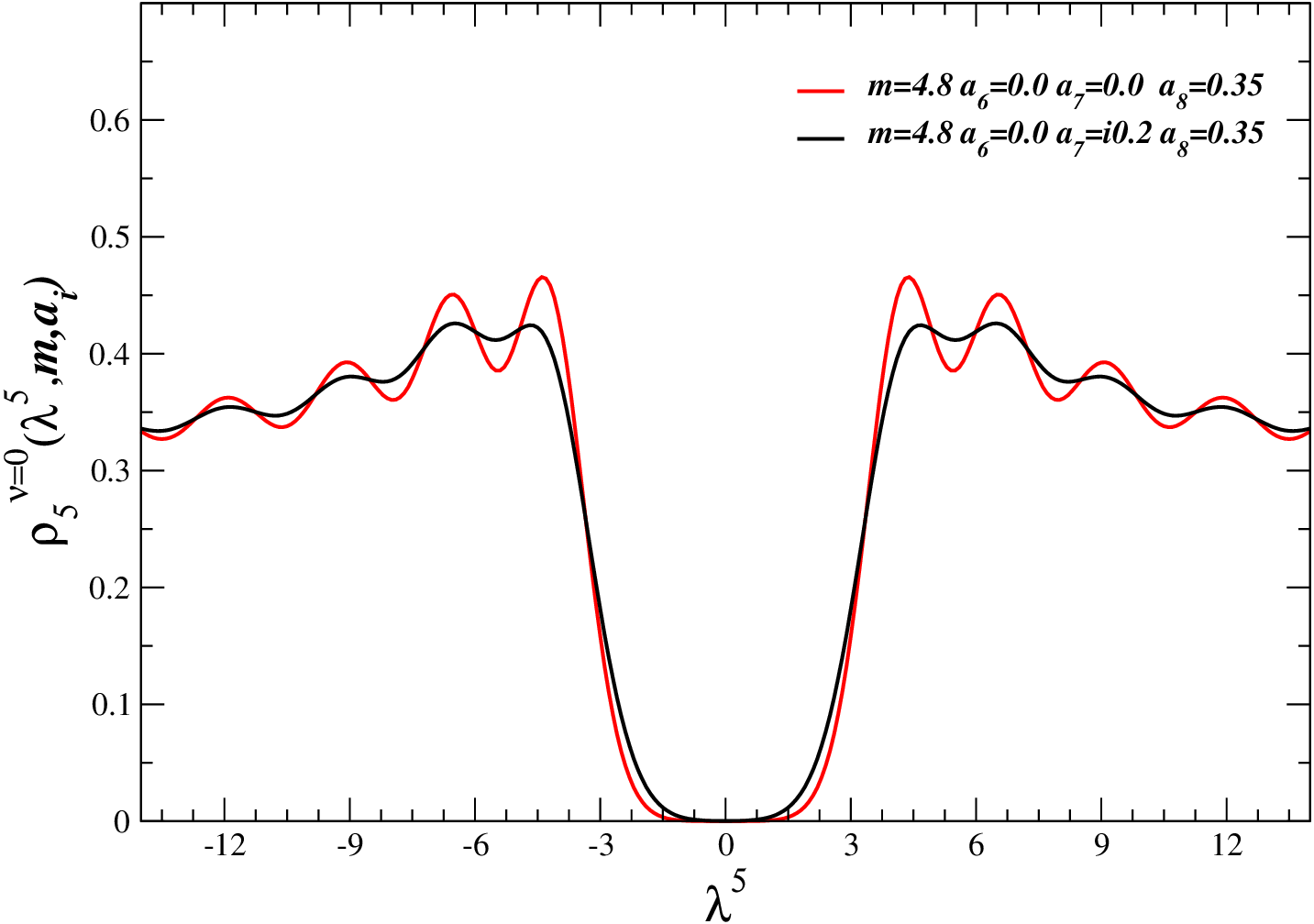}
\caption{\label{fig:a6a7dep} 
The eigenvalue density of $D_5$ for $\nu=0$ {\bf top:} the
dependence on $\ha_6$ and {\bf bottom:} the dependence on $\ha_7$. The
more oscillating (Red) line in both plots has $a_6=a_7=0$ and corresponds to 
the (Red) analytic curve which is plotted 
against the data in the top panel of Figure \ref{fig:fit2largemass}. Both 
$\ha_6^2<0$ or $\ha_7^2<0$ tends to smear out the
oscillations. Since the magnitude of the oscillations in the lattice
data presented in Figure \ref{fig:fit2largemass} is of the same
magnitude as those in the analytic prediction with $\ha_6=\ha_7=0$ we
can put an upper bound on these $|\ha_6^2|,|\ha_7^2|<0.01$. }
\end{figure}
\end{center}

\subsection{Individual eigenvalue distributions}

In order to understand in greater detail whether our lattice data match the
predictions from WCPT and WRMT it is advantageous to consider the
distributions of individual eigenvalues. As we do not have any simple
expression available from the chiral Lagrangian approach, we will
simply make use of the relation to Wilson chiral random matrix theory,
and generate the distributions numerically. See Figure \ref{fig:1st2nd}.

For the $\beta_{Iw}=2.635$ data on the $16^4$ lattice the first
eigenvalue matches well that generated from WRMT. The second
eigenvalue however is somewhat broader.

In the $\beta_{Iw}=2.79$ data from the $20^4$ lattice the first
eigenvalue again matches well that generated from WRMT. This time 
the second eigenvalue is repelled from the first in a manner
comparable to the one observed in the simulation of WRMT.

\begin{center}
\begin{figure}[t!]
\includegraphics[width=8cm,angle=0]{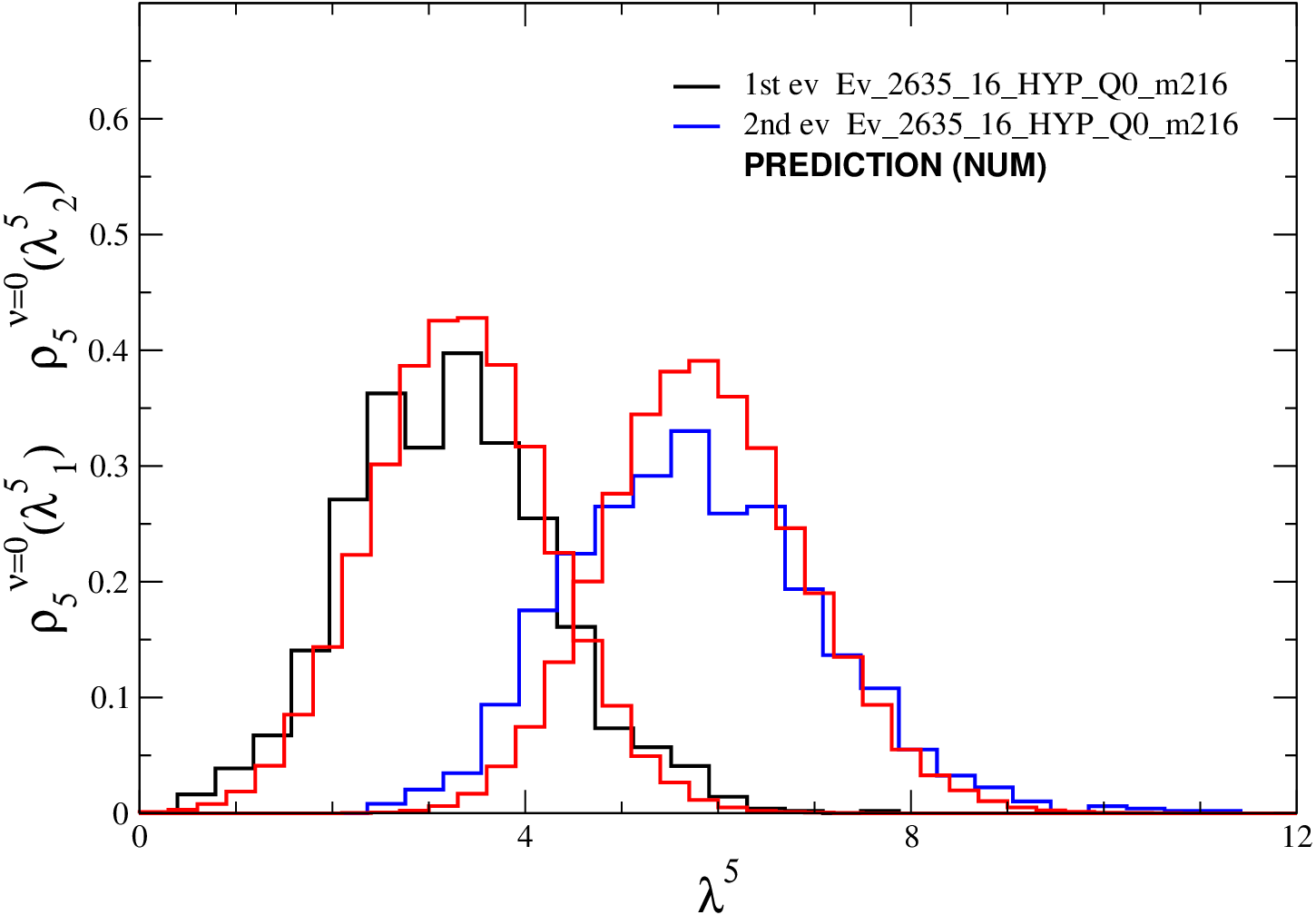}
\vspace{10mm}
 \vfill
\includegraphics[width=8cm,angle=0]{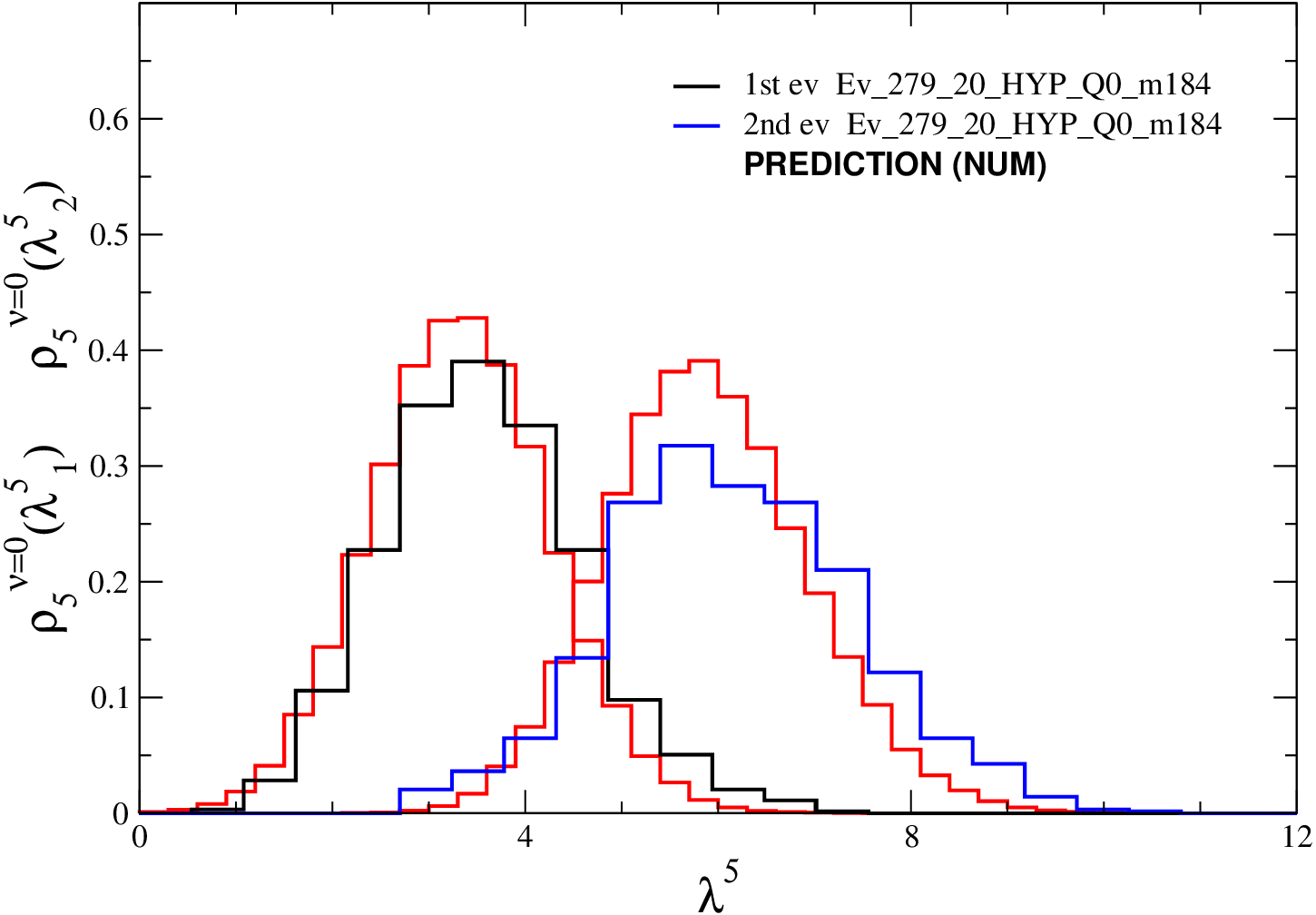}
\caption{\label{fig:1st2nd} 
  The distribution of the first ({\bf black}) and second ({\bf 
  blue}) eigenvalue {\bf top:} in the $16^4$ $\beta_{Iw}=2.635$ data set 
  and {\bf bottom:} in the $20^4$ $\beta_{Iw}=2.79$ data set. The
  ({\bf red}) histograms show the first and second eigenvalue
  distribution in a WRMT simulation with $n=100$ and 100000 matrices.}
\end{figure}
\end{center}

\subsection{The real eigenvalues of $D_W$}

From the flow of $\lambda^5$ as a function of $m$ we have also determined 
the real eigenvalues of $D_W$ as well as their chiralities. This enables us 
to measure both the density of the real modes
\be
\rho_{real}^{\nu}(\lambda^W;a) = \left\langle \sum_k \delta(\lambda^W-\lambda^W_k)\right\rangle
\ee  
as well as the distribution
\be
\rho_\chi^\nu(\lambda^W;a) = \left\langle \sum_k 
\delta(\lambda^W-\lambda^W_k){\rm sign}\langle k|\gamma_5|k\rangle\right\rangle.
\label{def:rhochi}
\ee 
The analytic predictions for both these quantities are available, see \cite{KVZ} for 
$\rho_{real}$ and \cite{DSV,ADSV} for $\rho_\chi$. The two distributions both merge to a 
$\nu$-fold $\delta$-function at the origin for $\hat{a}\to0$.

From the fit to $\rho_5$ we have already obtained the relevant values of $\Sigma V$, $\hat{m}$ 
and $\hat{a}$. This enables a highly nontrivial parameter free test of WCPT. In Figure 
\ref{fig:rhorealandchi} the parameter free analytic predictions are plotted along with the 
results from the lattice. Since accidentally the value $\hat{a}=0.35$ was obtained for both 
data sets we have combined the plots. For the plot we have rescaled the lattice eigenvalues, 
shown in Figure \ref{fig:rhorealLAT}, by $\Sigma V$ 
(the value for which we obtained from the respective fits to $\rho_5$) and then shifted the eigenvalues by 
$\Sigma V m_0 - \hat{m}$. For the $20^4$ data with $m_0=-0.184$ and $\beta_{Iw}=2.79$ this reads 
$\Sigma V m_0 - \hat{m} = 216(-0.184)-3.5 = -43.2$ while for the $16^4$ data with $\beta_{Iw}=2.635$ 
and $m_0=-0.216$ it reads $\Sigma V m_0 - \hat{m} = 157.5(-0.216)-3.5 =-37.5$. 

\begin{center}
\begin{figure}[t*]
\hspace{2mm}\includegraphics[width=8cm,angle=0]{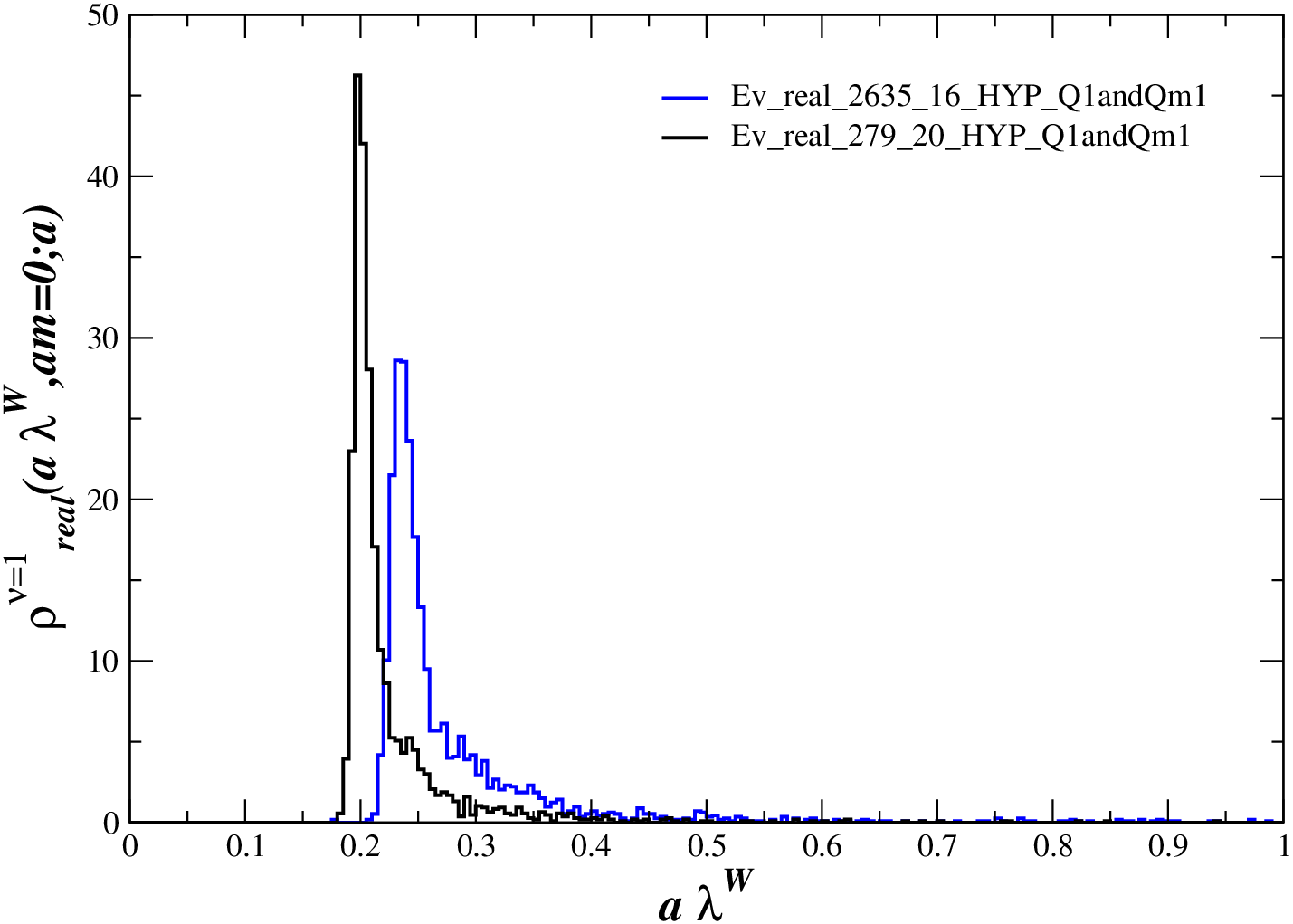}
\caption{\label{fig:rhorealLAT}
The distribution of the real modes of $D_W$ before subtraction of the bare mass and scaling of the axis.} 
\end{figure}
\end{center}

We observe that the scaling implied by WCPT is remarkable accurate. After the parameter free shift 
of the eigenvalues the center of the data peak is right at the origin as predicted by WCPT. Moreover, 
the width of the peak in both data sets is exactly of the right magnitude.
We also note that the peak height in the data increases towards the analytic prediction when going 
from $16^4$ to $20^4$. 

However, in the data there is an asymmetry of $\rho_{real}$ and $\rho_\chi$ which persist when going
from $16^4$ to $20^4$. Moreover, the average number of real modes is predicted to be 1.017 by WRMT
for $\nu=1$ and $\hat{a}=0.35$ \cite{KVZ}, in the data we find 1.237 for the $16^4$ lattice and 
1.185 for the $20^4$ lattice. That is, the average number of {\sl additional} real modes is 
off by a factor of 10. We note, however, that this average number of additional real modes decrease 
by $22\%$ when going from $a = 0.93$ fm on $16^4$ to $a = 0.075$ fm on $20^4$.

The asymmetry of the distribution of the real eigenvalues is a clear signal that at the present volumes 
there are small real eigenvalues of $D_W$ which are not captured by leading order WCPT. In \cite{Anna} 
it was observed that the eigenvectors corresponding to eigenvalues of $D_W$ which fall to the right of 
the main band have a strong tendency to be localized. Obviously such localized modes are not described 
by WCPT. 

\begin{center}
\begin{figure}[t*]
\hspace{2mm}\includegraphics[width=8cm,angle=0]{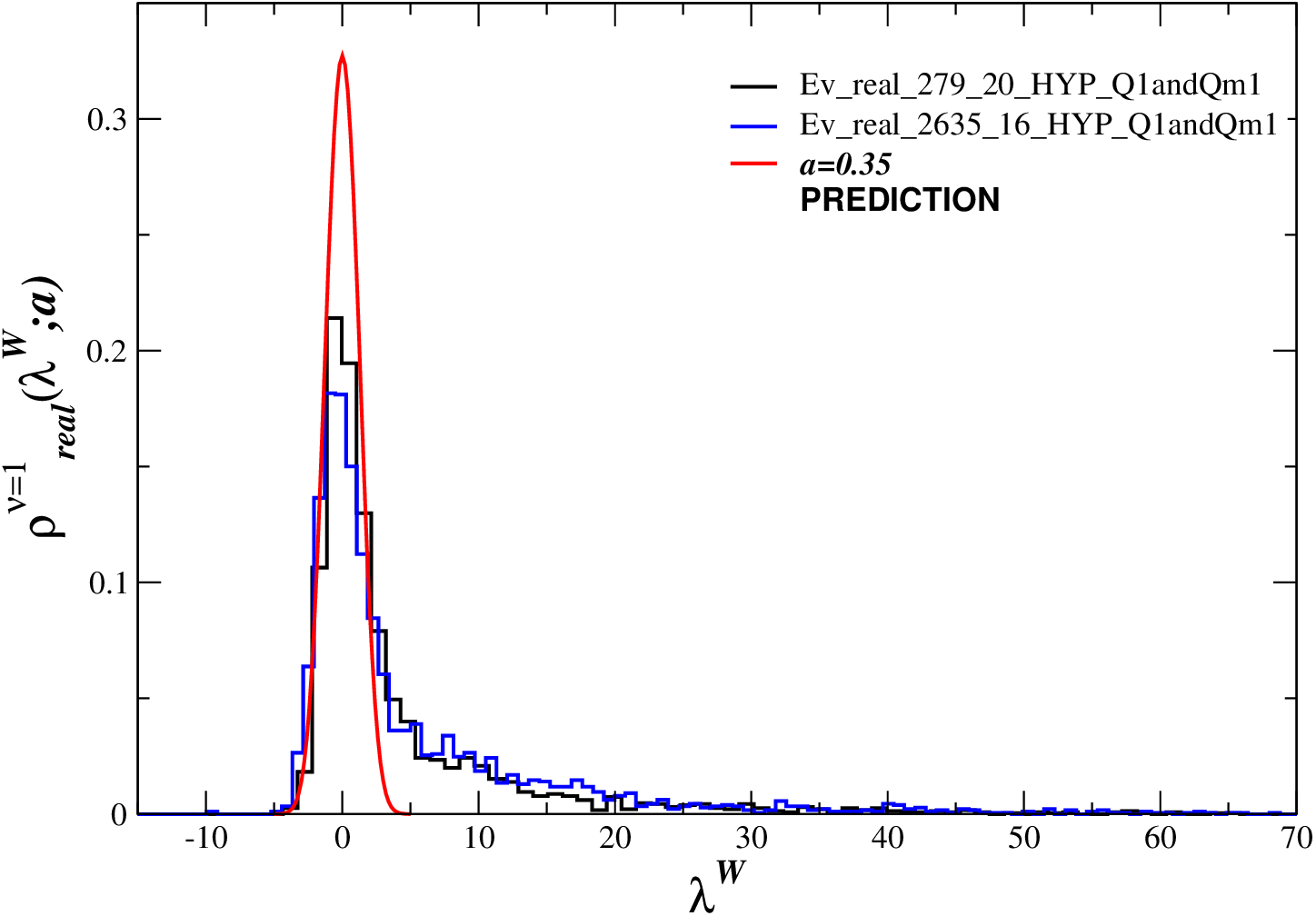}
\vspace{10mm}

\includegraphics[width=8cm,angle=0]{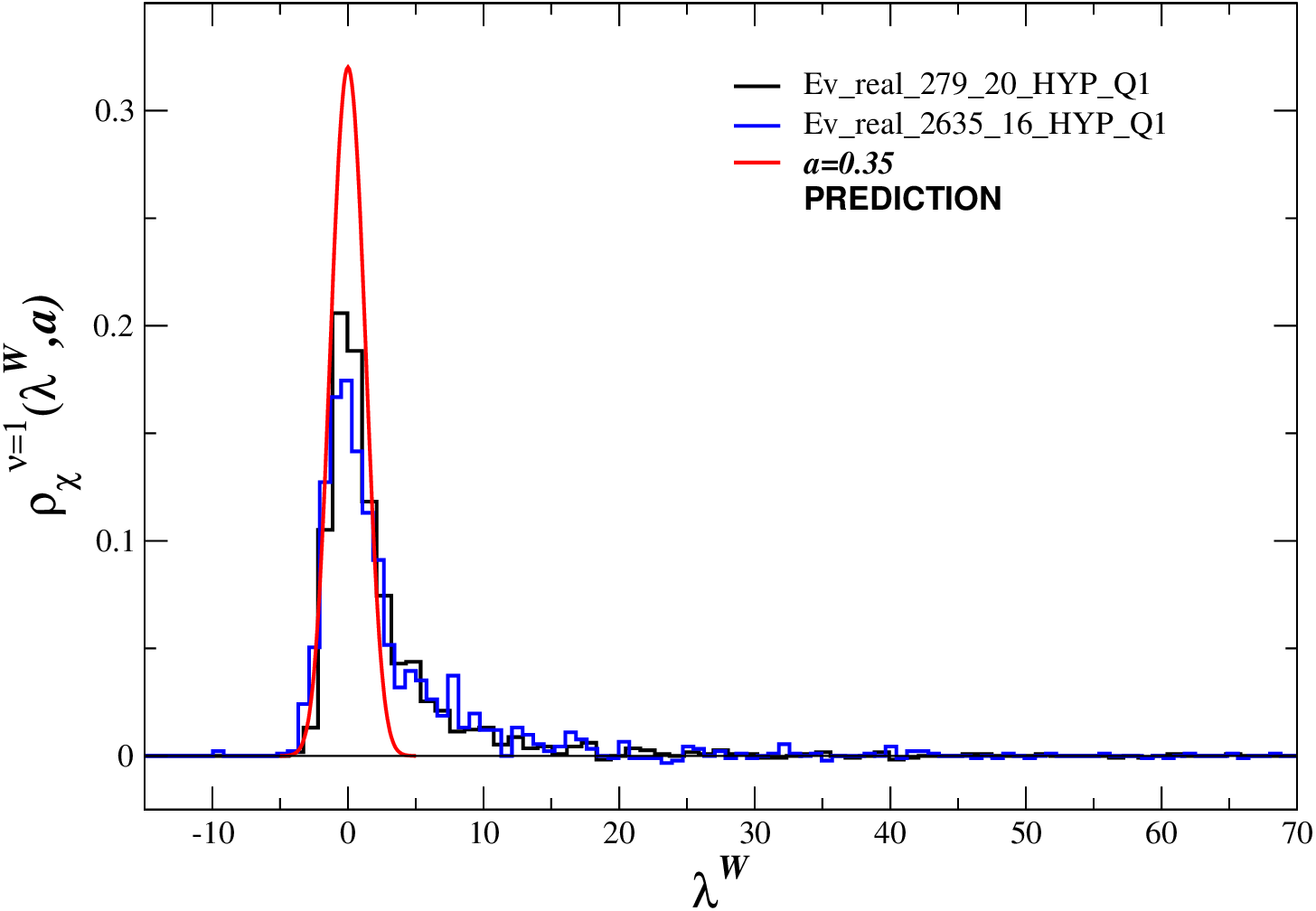}
\caption{\label{fig:rhorealandchi}
{\bf top:} The distribution of the real modes. {\bf bottom:} the distribution 
$\rho_\chi$ defined in Eq.~(\ref{def:rhochi}). All parameters are fixed from the fits to $\rho_5$. 
The eigenvalue data are first rescaled by $\Sigma V$ and then shifted by $m_0\Sigma V-\hat{m}$ 
where $m_0$ is the lattice mass and $\hat{m}$ is the value obtained previously for this $m_0$.} 
\end{figure}
\end{center}

\section{Conclusions}
\label{sec:conc}

The analytical predictions for the quenched microscopic spectral density of the
Hermitian Wilson-Dirac operator have been compared to the low lying eigenvalues
of Wilson fermions on the lattice. In this first study, we have
demonstrated that the results given for fixed index of the Wilson-Dirac 
operator can be matched by the predictions from Wilson chiral perturbation
theory with $W_8>0$ and $W_6=W_7=0$. We have successfully tested the finite 
volume scaling at fixed lattice spacing as well as the mass scaling as fixed 
lattice spacing predicted by Wilson chiral perturbation theory against the 
lattice data.  The spectrum of $D_5$ in this way gives
a very direct way of measuring the low-energy constants of Wilson
chiral perturbation theory. Although these constants represent
lattice artifacts, it is essential to know their values in order to
extract the physical constants. Moreover, the detailed predictions can 
be used with advantage to understand how the spectrum can be made to 
retain its gap as simulations are pushed towards physical values and
towards the continuum limit.

We emphasize that the assignment of index $\nu$ to a given gauge field
configuration must be based on the spectral flow of the Hermitian
Wilson-Dirac in order for the match to Wilson chiral perturbation
theory to be valid. In the microscopic limit the analytic results 
predict that the real eigenvalues of $D_W$ are located in well
separated regions. On the lattice, it should therefore be possible 
in a clean way to introduce a cut-off such that we consider only the 
physical branch of the spectrum. In support of this we have explicitly 
computed the real eigenvalues of $D_W$ and found that the bulk of the 
real eigenvalues follow the predictions from WCPT. However, we observed 
also an asymmetry of the distribution of the real eigenvalues not 
described by WCPT. For this reason some configurations that are difficult 
to classify remain in our samples. Similar concerns have been voiced 
already in the context of
the overlap operator \cite{Anna}, where localized modes that intuitively
should play no role in the continuum limit nevertheless significantly
deform the spectrum of the overlap operator at fixed $\nu$. We stress
that such localized modes are not described by Wilson chiral
perturbation theory. 
As such modes are more easily identified in the complex part of the
spectrum of $D_W$ a careful and high-statistics study of the
microscopic spectrum of the Wilson-Dirac operator which address this
issue would provide a valuable test of Wilson chiral perturbation
theory at presently realistic lattice spacings. The analytic
prediction for the quenched microscopic eigenvalue density of $D_W$ in 
the complex plane has been presented very recently \cite{KVZ}.

It is also possible to transform all analytical predictions for the
microscopic spectral density  of the Hermitian Wilson-Dirac operator
into corresponding expressions for the pseudoscalar condensate $\langle
\bar{\psi}\gamma_5\psi\rangle$, as discussed in Ref.~\cite{Heller}. 
It would be interesting to use this quantity to extract the Wilson 
low-energy constants.

To summarize we have demonstrated here that the theoretical predictions 
of Wilson chiral perturbation theory with $W_6=W_7=0$ and $W_8>0$ for 
the microscopic spectral properties of the Wilson-Dirac operator can 
be matched to lattice data. It would be 
most interesting to extend this quenched study to the physical case 
of two light flavors. The detailed theoretical predictions 
have been given recently \cite{SVdyn}.

\noindent
{\bf End note:}
We learned at the Lattice 2011 meeting that Albert Deuzeman, Urs Wenger 
and Ja\"ir Wuilloud have been investigating the same issues we
have discussed here. We thank these authors for coordinating
publication of results. Their paper is Ref.~\cite{DWW}.

\noindent
{\bf Acknowledgments:}
We would like to thank Jac Verbaarschot, Anna Hasenfratz, Hidenori Fukaya, 
Christian Lang and Christof Gattringer for discussions. 
The work of K.S. was supported by the {\sl Sapere Aude} program of
The Danish Council for Independent Research.


\end{document}